\documentclass[a4paper,11pt]{article}

\usepackage{jcappub} 

\usepackage[T1]{fontenc} 
\usepackage{amsmath}
\usepackage{amssymb}
\usepackage{graphicx} 

\usepackage{array}
\usepackage{xcolor}
\usepackage{cancel}
\usepackage[utf8x]{inputenc}
\usepackage[left]{lineno}
\usepackage[normalem]{ulem}
\usepackage{lineno}
\usepackage{orcidlink}

\title{Bayesian exploration of weakly interacting models of dark matter admixed neutron stars based on multi-density constraints}

\author{Payaswinee Arvikar\,\orcidlink{0009-0008-2007-9859}\textsuperscript{1,2}}
\author{Prashant Thakur\,\orcidlink{0000-0003-4189-6176}\textsuperscript{3}}
\author{Sarmistha Banik\,\orcidlink{0000-0003-0221-3651} \textsuperscript{2}}

\affiliation{\textsuperscript{1}Dharampeth M. P. Deo Memorial Science College, Nagpur 440033, India}
\affiliation{\textsuperscript{2}Department of Physics, BITS-Pilani Hyderabad Campus, Hyderabad 500078, India}
\affiliation{\textsuperscript{3}Department of Physics, Yonsei University, Seoul 03722, South Korea}

\emailAdd{p20230533@hyderabad.bits-pilani.ac.in}
\emailAdd{prashant@yonsei.ac.kr}
\emailAdd{sarmistha.banik@hyderabad.bits-pilani.ac.in}

\abstract{
We present a unified Bayesian analysis of Higgs-portal,
neutron-decay, and axion-like-particle (ALP)-mediated dark matter
in neutron stars within the single-fluid approximation. Using a
common relativistic mean-field description of hadronic matter,
we combine finite-nucleus and heavy-ion collision data, chiral
effective field theory, NICER measurements of four pulsars, and
GW170817. The individual dark-sector parameters remain weakly
determined, while model-dependent shifts in nuclear skewness
reveal the sensitivity of the inferred hadronic equation of state
to the assumed dark-matter mechanism. The three scenarios yield
substantially overlapping mass--radius and tidal-deformability
predictions, with median maximum masses near $2.1\,M_\odot$.
For $1.4\,M_\odot$ stars, the Higgs-portal model predicts higher
$f$-mode frequencies and shorter damping times than the
neutron-decay and ALP models, although their credible regions
overlap. Comparison with three consistently reanalysed two-fluid
models gives the largest Bayesian evidence to neutron decay:
under the adopted priors, its Bayes factors relative to the
Higgs-portal and two-fluid alternatives are approximately
$24$--$37$, but only $2.2$ relative to ALP. This ranking concerns
complete dark-matter models and does not establish a general
preference for either fluid description or evidence for a dark
component itself. Improved nuclear constraints and future
gravitational-wave asteroseismology could provide complementary
information to disentangle the hadronic and dark-sector
contributions.
}
\begin{document}
\maketitle

\section{Introduction}

Neutron stars (NSs) are compact remnants of massive stars, containing matter at densities several times larger than the nuclear saturation density. The star core is mainly composed of hadronic matter (HM), however the composition and behavior of matter under these conditions remain uncertain because such high densities are beyond the range of terrestrial laboratories. The equation of state (EOS), which relates the pressure and energy density, describes the microscopic composition of matter that is used to study macroscopic NS properties such as the mass, radius, compactness, and tidal deformability. A realistic EOS must therefore be consistent with both terrestrial nuclear experiments and astrophysical observations. Relativistic mean-field (RMF) models, in which nucleon--nucleon interactions are mediated by scalar and vector mesons together with meson self- and cross-interactions, have been widely employed to describe finite nuclei and dense stellar matter \cite{Dutra_2014}. 
The model parameters are fixed from nuclear physics experiments. 
The increasing precision of radio pulsar masses, NICER mass--radius measurements, and tidal-deformability constraints from binary mergers such as GW170817 provides an opportunity to constrain these scenarios.

On much larger scales, observations of galaxy rotation curves, gravitational lensing, galaxy clusters, the cosmic microwave background, and large-scale structure require a non-luminous matter component. Cosmological observations indicate that baryonic matter contributes approximately \(5\%\) of the total mass--energy content of the Universe, whereas nearly \(27\%\) is attributed to dark matter (DM). Its microscopic nature is, however, still unknown despite direct searches such as XENON100 \cite{Aprile_2012}, PandaX-II \cite{Wang_2020}, XENONnT \cite{XENON:2024wpa} and LUX-ZEPLIN \cite{LZ:2019sgr}, indirect-detection experiments \cite{DONATO2014,Heros2020,Biondini:2023}, and collider searches. The strong gravitational potential and large baryonic density of NSs make them promising laboratories for connecting this cosmological evidence to particle physics. DM may be inherited during stellar formation, captured from the Galactic halo through scattering with baryons, accreted during the lifetime of the star, or produced inside the star through non-standard particle processes. If a sufficiently large DM population is retained, it can modify the stellar EOS and leave observable imprints on the global properties of the star.

The unknown nature of DM has led to speculation of a wide range of candidates, including WIMPs, sterile neutrinos, neutralinos, axions, and other fermionic or bosonic particles. Fermionic DM is supported by degeneracy pressure, with repulsive self-interactions further controlling the stiffness of its EOS, whereas bosonic configurations are governed mainly by condensate dynamics and self-interactions \cite{Ellis_2018,Li_2012,das2020dark,Kouvaris_2010,Karkevandi2021}. Depending on the assumed interaction between DM and HM, DM-admixed NSs are generally described using single-fluid or two-fluid approaches. In the two-fluid framework, HM and DM interact only through gravity and their energy--momentum are conserved separately, allowing the formation of either a dark core or an extended halo depending on the DM mass, interaction strength, and abundance \cite{Karkevandi_2024,Rutherford_2023,thakur2023exploring,Arvikar:2025,Arvikar:2026,Miao_2022,AnkitKumar_2024,Shawqi:2024}. Quasi-static two-fluid calculations of protoneutron stars also show that DM cores can raise the equilibrium baryonic temperature through gravitational compression, whereas sufficiently large DM halo fractions can lower it
\cite{Issifu:2025jac}.  In the single-fluid framework, the two sectors are connected through non-gravitational interactions or chemical-equilibrium conditions and are described by an  EOS used to solve Tolman--Oppenheimer--Volkoff (TOV) equations. Several microscopic mechanisms have been proposed within this framework. Fermionic DM may interact with nucleons through the Higgs portal, where Higgs-boson exchange modifies the effective masses and the total energy density and pressure \cite{Das_2019,Routaray_2023,Flores:2024hts,Sen:2024yim}. Alternatively, the interaction may be mediated by an axion-like particle (ALP), providing a pseudoscalar portal between the dark and hadronic sectors \cite{Klangburam:2025,PThakur_2026}. A different possibility arises from the neutron-lifetime anomaly, for which an exotic decay channel such as \(n\rightarrow\chi+\phi\) produces a nearly degenerate dark fermion \(\chi\) and a light boson $\phi$; chemical equilibrium with neutrons then determines its abundance, while repulsive dark-sector interactions provide additional pressure support \cite{Shirke:2024ymc,Thakur_2025}. Although these three mechanisms have different particle-physics origins, they can lead to similar changes in NS observables. Increasing the dark component generally softens the effective EOS and reduces the maximum mass, radius, and tidal deformability, whereas sufficiently strong repulsive interactions can counteract this softening and support massive NSs \cite{Xiang_2014,Ellis_2018,das2020dark,Grippa2025}.

Beyond static observables, the non-radial oscillation modes of NSs provide a complementary, dynamical probe of their interior composition. Among these, the fundamental ($f$) mode couples most strongly to gravitational radiation: its frequency, typically in the range $1$--$3$~kHz, scales with the square root of the mean stellar density, while its damping time is set by gravitational-wave (GW) emission, making it a prime target of GW asteroseismology \cite{Andersson:1997rn}. Since an admixed dark component modifies the stellar density profile, it leaves a systematic imprint on the $f$-mode characteristics, as demonstrated for different DM models in both single-fluid and two-fluid treatments \cite{Shirke:2024ymc,Thakur:2025fmode,Chandrakar:2026fmode,Shahrbaf:2026fmode,Rather:2025fmode}. The $f$-mode frequency and damping time therefore constitute an additional observable through which the three scenarios considered in this paper may eventually be distinguished by next-generation GW detectors.

However, the effects of the dark sector are strongly degenerate
with uncertainties in the hadronic EOS, and different DM mechanisms
can predict similar mass--radius and tidal-deformability relations.
Bayesian inference may consistently combine nuclear and
astrophysical information, propagating uncertainties, and
determining correlations among the hadronic and dark-sector
parameters. Previous Bayesian and statistical studies have
investigated individual fermionic or bosonic DM models in either
single-fluid or two-fluid configurations, with the inferred DM
content ranging from a few percent to several tens of percent
depending on the particle model, priors, hadronic EOS, and definition
of the DM fraction
\cite{Husain_2021,Karkevandi2021,Routaray_2023,thakur2023exploring,
Rutherford_2023,Karkevandi_2024,Arvikar:2025,Arvikar:2026}.
These results cannot be compared directly because the different
microscopic mechanisms have generally been studied using different
nuclear models, parameter ranges, and observational constraints.
To the best of our knowledge, a unified Bayesian analysis of the
Higgs-portal, axion-like-particle-mediated, and neutron-decay
scenarios within the single-fluid framework is yet to be performed.
In the present work, we place these three models on the same
statistical footing by employing a common hadronic framework and
a consistent set of nuclear-physics and multimessenger astrophysical
constraints. We infer the posterior distributions of the relevant
dark-sector parameters, propagate their uncertainties to the EOS
and NS observables, and investigate whether present data can
distinguish the three interaction mechanisms. A central aim of
this work is to determine whether current data favor particular
microscopic DM scenarios. We address this through Bayesian evidence
comparisons of the three single-fluid models and three reanalysed
two-fluid models, using common nuclear and astrophysical
constraints and accounting for uncertainties in both sectors
under the adopted priors.

The paper is organised as follows. In section \ref{method}, we describe the HM and DM models and the corresponding EOSs. Particularly, sub-sections \ref{sec:structure},\ref{constraints} and \ref{bayes_inf} explain the methodology for the analyses. In section \ref{results}, we discuss our results and we conclude the findings in section \ref{conclusion}.

\section{Methodology}\label{method}

Hadronic matter is described within the RMF approximation
(Sec.~\ref{sec:hm}), and the dark sector is realized through three
mechanisms: (i) a fermionic DM particle coupled to nucleons through
the Higgs portal, (ii) a dark fermion produced through an exotic
neutron-decay channel, and (iii) a fermionic DM particle coupled to
nucleons through a light pseudoscalar ALP. In all three cases the two
sectors are connected by non-gravitational interactions, and
DM-admixed NS matter forms a single fluid,
\[
{\mathcal{L} = \mathcal{L}_{\rm HM} + \mathcal{L}_{\rm DM}, \qquad
\mathcal{E} = \mathcal{E}_{\rm HM} + \mathcal{E}_{\rm DM}, \qquad
P = P_{\rm HM} + P_{\rm DM},}
\]
with additive energy density and pressure in the mean-field
approximation (Sec.~\ref{sec:dm}). The EOS of each model is used in
the TOV and tidal-deformability calculations
(Sec.~\ref{sec:structure}). The parameters of all three models are
constrained by the same nuclear and astrophysical data through
Bayesian inference (Secs.~\ref{constraints} and \ref{bayes_inf}).

\subsection{Hadronic matter}\label{sec:hm}

Nucleon--nucleon interactions are mediated by the scalar--isoscalar meson $\sigma$, the vector--isoscalar meson $\omega$, and the vector--isovector meson $\varrho$, supplemented by non-linear self- and cross-couplings that control the high-density behavior of the EOS. The Lagrangian density is \cite{Dutra_2014,Venneti_2024},
\begin{eqnarray}\label{eq:lag}
    \nonumber
   \mathcal{L}_{\text{HM}}&=& \!\bar{\psi}\left(i\gamma^\mu\partial_\mu - M\right)\psi +\!\!g_\sigma \sigma \bar{\psi} \psi -\!\!g_\omega \bar{\psi}\gamma^\mu\omega_\mu\psi -\frac{g_\rho}{2}\bar{\psi}\gamma^\mu{\vec{\tau}\cdot}\vec{\rho}_\mu{\psi} +\frac{1}{2}\left(\partial^\mu\sigma\partial_\mu\sigma - m^2_\sigma \sigma^2\right) \\ \nonumber
   &&- \frac{\text{A}}{3} \sigma^3 - \frac{\text{B}}{4} \sigma^4 \! -\frac{1}{4} \Omega^{\mu\nu}\Omega_{\mu\nu} + \frac{1}{2}m^2_\omega\omega^\mu\omega_\mu +\frac{\text{C}}{4}\left(g_\omega^2 \omega_\mu \omega^\mu\right)^2 \\
   &&-\frac{1}{4}\vec{B}^{\mu\nu}\vec{B}_{\mu\nu} + \frac{1}{2}m^2_\rho \vec{\rho}_\mu \vec{\rho}^{ \mu} + \frac{1}{2} \Lambda_v g^2_\omega g^2_\rho \omega_\mu \omega^\mu \vec{\rho}_\mu \vec{\rho}^\mu,
\end{eqnarray}
where $\Omega_{\mu\nu} = \partial_\nu \omega_\mu - \partial_\mu \omega_\nu$ and $\vec{B}_{\mu\nu} = \partial_\nu \vec{\rho}_\mu - \partial_\mu \vec{\rho}_\nu - g_\rho (\vec{\rho}_\mu \times \vec{\rho}_\nu)$ are the vector-field tensors and $\vec{\tau}$ is the nucleon isospin operator. In the mean-field approximation the meson fields take their ground-state expectation values, and the energy density and pressure of nucleonic matter read \cite{Venneti_2024}
\begin{eqnarray}{\label{eos_nm}}
   \mathcal{E}_{HM} = \frac{1}{\pi^2} \int_{0}^{k_p,k_n} dk k^2 \sqrt{k^2 + (M^*)^2}  + \frac{1}{2} m_\sigma^2 \sigma^2 - \frac{1}{2} m_\omega^2 \omega^2 -  \frac{1}{2} m_\rho^2 \rho^2 + \frac{\text{A}}{3} \sigma^3 + \frac{\text{B}}{4} \sigma^4 \nonumber \\ \nonumber
   - \frac{\text{C}}{4} g_\omega^4 \omega^4 + g_\omega \omega(\rho_p + \rho_n) - \frac{\Lambda_v}{2} (g_\rho g_\omega \rho \omega )^2 + \frac{g_\rho}{2} \rho (\rho_p - \rho_n), \\ \nonumber
   P_{HM} = \frac{1}{3 \pi^2} \int_{0}^{k_p,k_n} dk \frac{k^4} {\sqrt{k^2 + (M^*)^2}} - \frac{1}{2} m_\sigma^2 \sigma^2 + \frac{1}{2} m_\omega^2 \omega^2 +  \frac{1}{2} m_\rho^2 \rho^2 - \frac{\text{A}}{3} \sigma^3 - \frac{\text{B}}{4} \sigma^4 \\ 
   + \frac{\text{C}}{4} g_\omega^4 \omega^4 +  \frac{\Lambda_v}{2} (g_\rho g_\omega \rho \omega )^2,
\end{eqnarray}
where $\rho_{p,n}$ and $k_{p,n}$ are the proton and neutron densities and Fermi momenta, respectively, $M^{*}=M-g_{\sigma}\sigma$ is the effective nucleon mass, and $\sigma$, $\omega$, $\rho$ denote the meson mean fields. Rather than sampling the couplings directly, we characterize the hadronic sector by seven nuclear matter parameters (NMPs) evaluated at saturation---the saturation density $\rho_0$, binding energy per nucleon $E_0$, incompressibility $K_0$, skewness $Q_0$, effective-mass ratio $M^*/M$, symmetry energy $J_0$, and its slope $L_0$---and map each sampled set onto the couplings $\{g_\sigma,g_\omega,g_\rho,A,B,C,\Lambda_v\}$ \cite{Venneti_2024}. The corresponding priors are listed in Table~\ref{tab:NMP_priors}.

\subsection{Dark matter models}\label{sec:dm}

We now specify the dark-sector Lagrangian
$\mathcal{L}_{\rm DM}$ and the corresponding energy density
$\mathcal{E}_{\rm DM}$ and pressure $P_{\rm DM}$ for each
of the three scenarios introduced above. The adopted
dark-sector parameter priors are listed in
Table~\ref{tab:DMP_priors}.

\subsubsection{Higgs portal}

In the first scenario, a fermionic DM particle $\chi$
couples to the Higgs field $h$ with strength $y$, while nucleons
couple to $h$ through the effective Yukawa coupling $fM/v$.
Here $M$ is the bare nucleon mass appearing in
Eq.~\ref{eq:lag}, $f=0.35$ is the nucleon--Higgs form factor, and
$v=250$~GeV is the Higgs vacuum expectation value. The
dark-sector Lagrangian, including the interaction with the
nucleon field $\psi$, is \cite{Das_2019,Flores:2024hts}
\begin{equation}
\begin{aligned}
\mathcal{L}_{\rm DM}
&= \bar{\chi}
   (i\gamma^\mu\partial_\mu-m_\chi+yh)\chi \\
&\quad + \frac{1}{2}\partial_\mu h\,\partial^\mu h
       - \frac{1}{2}m_h^2h^2
       + f\frac{M}{v}\bar{\psi}h\psi .
\end{aligned}
\end{equation}
Here $m_\chi$ is the DM particle mass and
$m_h=125$~GeV is the Higgs mass. In the mean-field
approximation, the DM effective mass is
$m_\chi^*=m_\chi-yh_0$, where $h_0$ denotes the Higgs mean
field. The DM kinetic and Higgs-field contributions to the
energy density and pressure are
\begin{align}
\mathcal{E}_{DM}
&= \frac{1}{\pi^2}
   \int_0^{k_F^{DM}}
   k^2\sqrt{k^2+(m_\chi^*)^2}\,dk
   + \frac{1}{2}m_h^2h_0^2,
   \nonumber \\
P_{DM}
&= \frac{1}{3\pi^2}
   \int_0^{k_F^{DM}}
   \frac{k^4}{\sqrt{k^2+(m_\chi^*)^2}}\,dk
   - \frac{1}{2}m_h^2h_0^2 .
\end{align}
The DM Fermi momentum $k_F^{DM}$ determines the number
density through $n_{DM}=(k_F^{DM})^3/(3\pi^2)$. For the Higgs-portal model, we vary three dark-sector
parameters in the Bayesian analysis: the DM mass $m_\chi$,
the DM--Higgs coupling $y$, and the DM Fermi momentum
$k_F^{DM}$.

\subsubsection{Neutron-decay anomaly}

The second scenario is motivated by the neutron-lifetime
anomaly: bottle experiments count surviving neutrons, whereas
beam experiments count protons produced in standard
$\beta$ decay. The measurements reported in
Refs.~\cite{Pattie:2017vsj,Yue:2013qrc} yield a shorter bottle
lifetime, with a discrepancy of approximately $4\sigma$.
An additional neutron decay channel into the dark sector
was proposed as a possible explanation of this difference~\cite{Fornal:2018eol}; here we consider
\begin{equation}\label{eqn:darkdecay}
n\rightarrow\chi+\phi~,
\end{equation}
where $\chi$ is a dark spin-1/2 fermion and $\phi$ a light dark boson. Inside a NS the light boson escapes the star, while the dark fermions remain gravitationally bound and equilibrate with the surrounding matter, so that the chemical-equilibrium condition
\begin{equation}\label{eqn:chemicalequilibrium}
    \mu_{\chi}=\mu_{n}
\end{equation}
fixes the local DM abundance. Nuclear stability requires $937.993 \, \text{MeV} < m_{\chi} + m_{\phi} < m_n = 939.565 \, \text{MeV}$ \cite{Shirke23}, and stability of the dark particles against further $\beta$-decay requires $|m_{\chi} - m_{\phi}| < m_p + m_e = 938.783 \, \text{MeV}$ \cite{Fornal:2020gto}; we set $m_\chi=938.0$~MeV. Since the conversion of neutrons alone would soften the EOS, repulsive DM self-interactions are introduced through a dark vector boson,
\begin{equation}
    \mathcal{L} \supset -g_V \bar{\chi} \gamma^{\mu} \chi V_{\mu} - \frac{1}{4} V_{\mu\nu} V^{\mu\nu} + \frac{1}{2} m_V^2 V_{\mu} V^{\mu}~,
\end{equation}
with coupling $g_V$ and mass $m_V$. The DM energy density and chemical potential then read
\begin{equation}
    \mathcal{E}_{DM} = \frac{1}{\pi^2} \int_{0}^{k_{F_{\chi}}} k^2 \sqrt{k^2 + m_{\chi}^2} \, dk + \frac{1}{2} G_{\chi} n_{\chi}^2,
    \label{eqn:endens_dm}
\end{equation}
\begin{equation} \label{eqn:Gdefinition}
    G_{\chi} = \left( \frac{g_V}{m_V} \right)^2, \qquad n_{\chi} = \frac{k_{F_{\chi}}^3}{3 \pi^2}, \qquad \mu_\chi = \sqrt{k_{F_\chi}^2 + m_\chi^2} + G_{\chi} n_\chi,
\end{equation}
and the pressure follows from the thermodynamic relation
\begin{equation}
P_{DM} = \mu_\chi n_\chi-\mathcal{E}_{DM}.
\end{equation}
The only dark-sector parameter sampled in this model is \(G_\chi\).

\subsubsection{ALP mediator}

The third scenario adopts the ALP-mediated model of our earlier work \cite{Klangburam:2025,PThakur_2026}: the Standard Model (SM) is extended by a Dirac fermion $\chi$ (the DM candidate) and a pseudoscalar ALP $a$ that mediates the DM--nucleon interaction,
\begin{align}
\mathcal{L}_{\rm DM}
&= \bar{\chi}(i\gamma^\mu\partial_\mu-m_\chi)\chi
 + \frac{1}{2}\partial_\mu a\,\partial^\mu a
 - \frac{1}{2}m_a^2a^2
 + \frac{m_\chi}{f_a}C_\chi
   \bar{\chi}i\gamma^5\chi\,a
 + \sum_f \left[
   \frac{m_f}{f_a}C_f
   \bar{\Psi}_f i\gamma^5\Psi_f\,a
   \right].
\end{align}
The effective couplings are defined as,
\begin{equation}
g_{aff} = \frac{m_f}{f_a}C_f,
\qquad
g_{a\chi\chi} = \frac{m_\chi}{f_a}C_\chi.
\end{equation}
Here $f_a$ is the ALP interaction scale, and $C_f$
and $C_\chi$ are dimensionless coefficients. We assume
universal $C_f$ for SM fermions, implying
$g_{aff}\propto m_f$. Production mechanisms and observational
constraints for related axion and ALP portal models are
discussed in
Refs.~\cite{Dror:2023fyd,Armando:2023zwz,Klangburam:2025,Yang_2024}.
In the mean-field approximation, the DM kinetic and ALP-field
contributions to the energy density and pressure are
\begin{align}
\mathcal{E}_{DM}
&= \frac{1}{\pi^2}
   \int_0^{k_F^{DM}}
   k^2\sqrt{k^2+(m_\chi^*)^2}\,dk
   + \frac{1}{2}m_a^2a_0^2,
   \nonumber \\
P_{DM}
&= \frac{1}{3\pi^2}
   \int_0^{k_F^{DM}}
   \frac{k^4}{\sqrt{k^2+(m_\chi^*)^2}}\,dk
   - \frac{1}{2}m_a^2a_0^2.
\end{align}
Here
$m_\chi^*=\sqrt{m_\chi^2+g_{a\chi\chi}^{\,2}a_0^2}$
is the DM effective mass, and $a_0$ is the ALP mean field,
determined self-consistently with the $\sigma$, $\omega$,
and $\rho$ fields \cite{Klangburam:2025}.
The two dark-sector parameters sampled in this analysis
are $m_\chi$ and $k_F^{DM}$.

\begin{table*}[t]
    \centering
    \renewcommand{\arraystretch}{2.5}
    \begin{tabular}{|c|c|c|c|c|c|c|}
        \hline
        $\rho_0$ (fm$^{-3})$ & E$_0$ (MeV)& K$_0$ (MeV)& Q$_0$ (MeV) & $M^*/M$ & J$_0$ (MeV)& L$_0$ (MeV)\\
        \hline
        0.14-0.17 & -16.5 - (-15.5) & 150 - 300 & -1500 - 400 & 0.5 - 0.8 & 20 - 40 & 20 - 100 \\
        \hline
    \end{tabular}
    \caption{Uniform priors defined for all HM parameters.} 
    \label{tab:NMP_priors}
\end{table*}

\begin{table*}[h]
    \centering
    \renewcommand{\arraystretch}{2.7}
    \begin{tabular}{|c|c|c|c|}
        \hline
        Model & Interaction parameter & $m_\chi$ (MeV) & $k_F^{DM}$ (MeV)\\
        \hline
        Higgs & y = 0.001 - 0.20 & log(100) - log(5$\times 10^5$) & 0 - 100 \\
        \hline
        N-decay & G$_\chi$=0.001-1000 & --- & --- \\
        \hline
        ALP & --- & log(100) - log(5$\times 10^5$) & 0 - 100 \\
        \hline
    \end{tabular}
    \caption{Uniform priors defined for all DM parameters.}
    \label{tab:DMP_priors}
\end{table*}

\subsection{Stellar structure}\label{sec:structure}

For each DM model, we determine the stellar structure
using the corresponding EOS of DM-admixed matter in the
TOV equations ($G=c=1$),
\begin{align}
\label{tov}
\frac{dP}{dr}
&= -(\mathcal{E}+P)
   \frac{m(r)+4\pi r^3P}{r[r-2m(r)]},
   \nonumber \\
\frac{dm}{dr}
&= 4\pi r^2\mathcal{E},
\end{align}
where $m(r)$ is the gravitational mass enclosed within
radius $r$. The tidal Love number $k_2$ is obtained by
solving the linear quadrupolar tidal-perturbation equations
on the stellar background \cite{Klangburam:2025}.
The dimensionless tidal deformability is then
$\Lambda=(2/3)k_2\mathcal{C}^{-5}$, where
$\mathcal{C}=M/R$ is the compactness, and $M$ and $R$
are the stellar mass and radius \cite{Arvikar:2025}.

\subsection{Constraints on the equation of state}
\label{constraints}

We constrain the three DM models using a common set of
theoretical, experimental, and astrophysical inputs, following
Ref.~\cite{Arvikar:2026}. At low densities, the hadronic EOS
is constrained by the pressures of $\beta$-equilibrated matter
from chiral effective field theory ($\chi$EFT)
\cite{Keller2023}, evaluated at baryon densities of
0.1, 0.15, and 0.2~fm$^{-3}$. Finite-nuclei (FNC) and
heavy-ion collision (HIC) data provide complementary
constraints on the symmetry energy $J(\rho)$, the pressure
of symmetric nuclear matter $P_{\rm SNM}(\rho)$, and the
symmetry pressure $P_{\rm sym}(\rho)$. The incompressibility
$K_0$ is constrained by the isoscalar giant monopole resonance.
The adopted nuclear constraints and their density ranges
are detailed in
Refs.~\cite{Tsang:2020,Venneti_2024,Arvikar:2026}.
Astrophysical constraints are supplied by the joint
mass--radius measurements of PSR J0030+0451
\cite{Riley_2019,Miller_2019}, PSR J0740+6620
\cite{Riley_2021,Miller_2021}, PSR J0437-4715
\cite{Choudhury:2024xbk}, and PSR J0614-3329
\cite{Mauviard:2025} from NICER, together with the
tidal-deformability constraint from GW170817
\cite{Abbott_2018}.

\subsection{Bayesian inference}\label{bayes_inf}

For each DM scenario, we infer the hadronic and
dark-sector parameters $\mathbf{\Theta}$ using Bayes'
theorem \cite{stuart1994kendalls},
$P(\mathbf{\Theta}|D)
=\mathcal{L}(D|\mathbf{\Theta})P(\mathbf{\Theta})/\mathcal{Z}$.
The prior distributions are specified in
Tables~\ref{tab:NMP_priors} and \ref{tab:DMP_priors}.
The Bayesian evidence $\mathcal{Z}$ normalizes the posterior
and is used for model comparison. Assuming conditional
independence between the four groups of constraints
described in Sec.~\ref{constraints}, given the model
parameters, we write
\begin{equation}\label{bayes}
\mathcal{L}(D|\mathbf{\Theta})
=\mathcal{L}_{\rm FNC/HIC}
\times \mathcal{L}_{\chi{\rm EFT}}
\times \mathcal{L}_{\rm NICER}
\times \mathcal{L}_{\rm GW}.
\end{equation}
The nuclear observables of Sec.~\ref{constraints} ($P_{SNM}$, $K_0$, $J(\rho)$, $P_{sym}(\rho)$; Table S3 of Ref.~\cite{Venneti_2024}) and the $\chi$EFT pressures enter through Gaussian likelihoods. The likelihoods of the four NICER pulsars are constructed from kernel density estimates of the full posterior samples, marginalized over a uniform central-mass distribution $m \in [1\,M_\odot,\, M_{max}(\mathbf{\Theta})]$, and the GW170817 likelihood is likewise marginalized over the component masses with $\Lambda_{1,2}(m,\mathbf{\Theta})$; the explicit expressions are identical to those of our earlier analyses \cite{Arvikar:2025,Arvikar:2026,Venneti_2024}. The posterior and evidence are computed with the sampler, \texttt{PyMultiNest} and considered 2000 live points. 

\section{Results and Discussions}\label{results}

\begin{figure*}
    \centering
    \includegraphics[width=\linewidth]{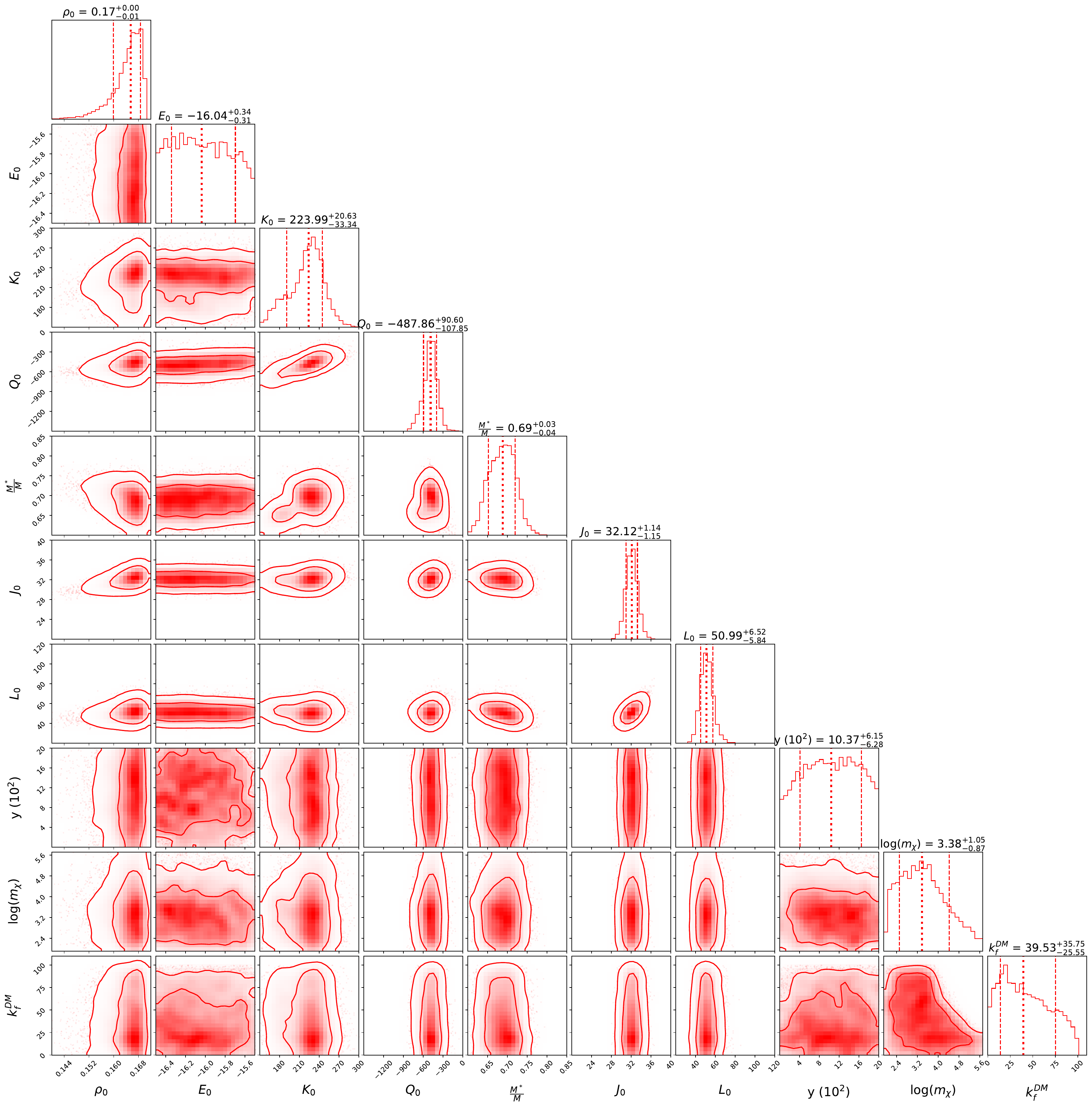}
    \caption{Marginalized one- and two-dimensional posterior distributions of
the nuclear-matter and dark-sector parameters for the Higgs-portal model,
obtained using the combined $\chi$EFT, nuclear-physics, NICER, and GW170817
constraints. Diagonal panels show the one-dimensional posteriors, with the
median indicated by the dotted line and the 68\% credible interval by the
dashed lines. Off-diagonal panels show the pairwise posteriors, with inner
and outer contours enclosing the 68\% and 95\% credible regions,
respectively. The
parameters are $\rho_0$, $E_0$, $K_0$, $Q_0$, $M^*/M$, $J_0$, $L_0$, $y$,
$m_\chi$, and $k_F^{\rm DM}$. The plotted dark-sector variables are
$10^2y$, $\log_{10}(m_\chi/{\rm MeV})$, and $k_F^{\rm DM}$ in MeV.
The quantities $E_0$, $K_0$, $Q_0$, $J_0$, and $L_0$ are given in MeV,
$\rho_0$ in fm$^{-3}$, and $M^*/M$ and $y$ are dimensionless.}
    \label{fig:corner_Higgs}
\end{figure*}

Figure~\ref{fig:corner_Higgs} presents the marginalized one- and
two-dimensional posterior distributions of the seven nuclear-matter
parameters and the three Higgs-portal dark-sector parameters. The combined
nuclear, $\chi$EFT, NICER, and GW170817 constraints produce comparatively
localized posteriors for several nuclear-matter parameters. The inferred
values at the 68\% credible level are
$K_0=223.99^{+20.63}_{-33.34}\,\mathrm{MeV}$,
$Q_0=-487.86^{+90.60}_{-107.85}\,\mathrm{MeV}$,
$M^*/M=0.69^{+0.03}_{-0.04}$,
$J_0=32.12^{+1.14}_{-1.15}\,\mathrm{MeV}$, and
$L_0=50.99^{+6.52}_{-5.84}\,\mathrm{MeV}$. In contrast, the binding
energy remains comparatively weakly constrained, with
$E_0=-16.04^{+0.34}_{-0.31}\,\mathrm{MeV}$. The posterior of the
saturation density is concentrated near the upper boundary of its adopted
prior, yielding $\rho_0=0.17^{+0.00}_{-0.01}\,\mathrm{fm}^{-3}$.
Consequently, the apparent localization of $\rho_0$ should be interpreted
with caution, as it may be sensitive to the chosen prior range.

The individual Higgs-portal parameters are less precisely determined.
Since the quantity displayed in the figure is $10^2y$, the corresponding
coupling is $y=0.104^{+0.062}_{-0.063}$. The dark-matter mass is constrained
to $\log_{10}(m_\chi/\mathrm{MeV})=3.38^{+1.05}_{-0.87}$, corresponding
to a median mass of approximately $2.4\,\mathrm{GeV}$ and a broad 68\%
credible range of approximately $0.32$--$26.9\,\mathrm{GeV}$. Similarly,
the inferred Fermi momentum,
$k_F^{\mathrm{DM}}=39.53^{+35.75}_{-25.55}\,\mathrm{MeV}$, remains
broadly distributed. These results indicate that the present data
constrain the hadronic sector more strongly than the individual
dark-sector parameters and do not select a sharply defined Higgs-portal
solution.

The two-dimensional posteriors show clear positive correlations between
$K_0$ and $Q_0$ and between $J_0$ and $L_0$, whereas the correlations
between the hadronic and dark-sector parameters are generally weak. A
broad degeneracy is visible between $m_\chi$ and $k_F^{\mathrm{DM}}$, with
large Fermi momenta preferentially allowed at lower dark-matter masses.
This behavior suggests that the observations primarily constrain the
combined effect of the dark component on the EOS rather than determining
$m_\chi$ and $k_F^{\mathrm{DM}}$ independently.

\begin{figure*}
    \centering
    \includegraphics[width=\linewidth]{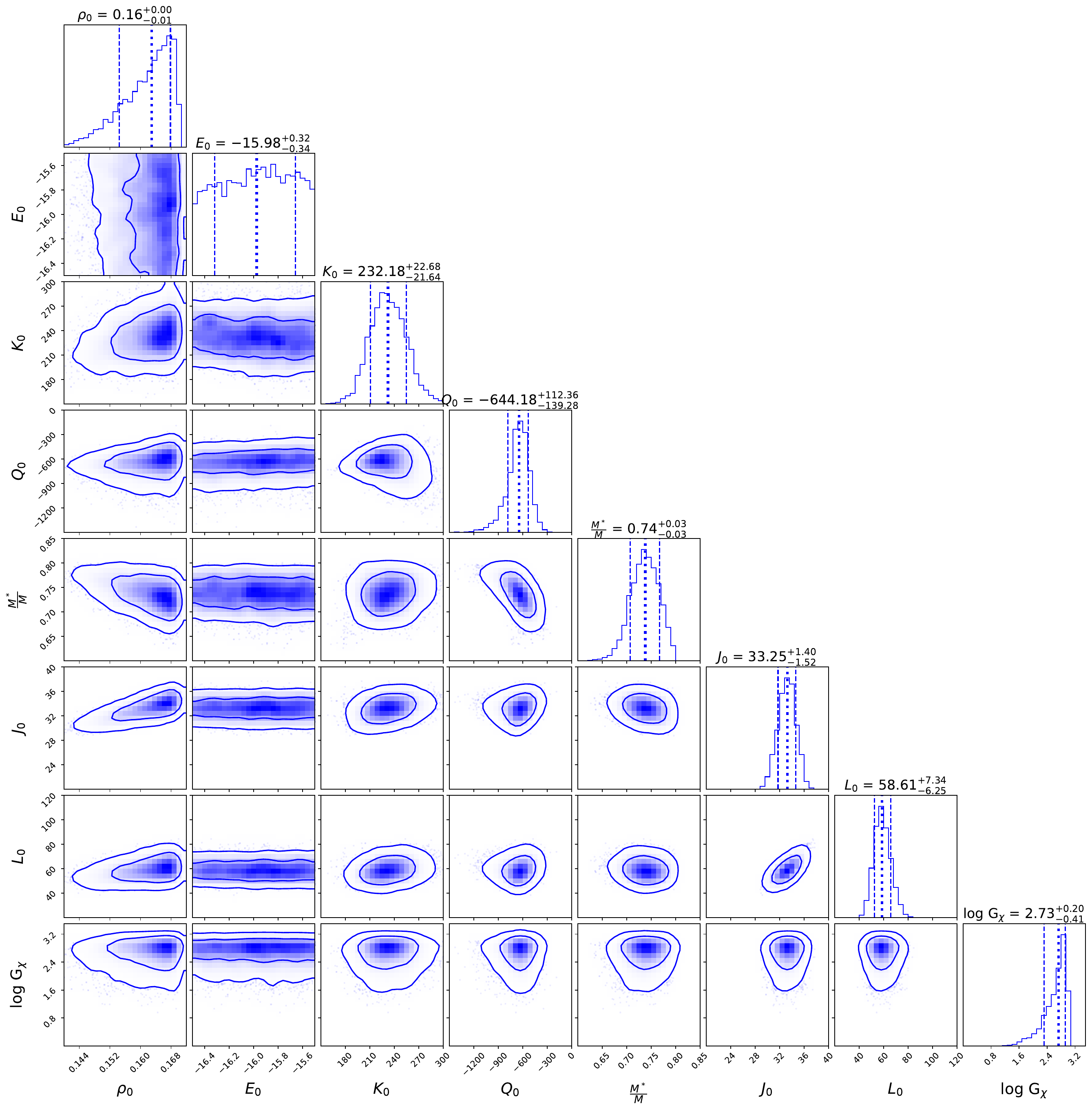}
    \caption{Same as Fig.\ref{fig:corner_Higgs} for Neutron decay mechanism, with the parameters $\rho_0, E_0, K_0, Q_0, \frac{M^*}{M},J_0,L_0, G_\chi$}
    \label{fig:corner_Ndecay}
\end{figure*}

Figure~\ref{fig:corner_Ndecay} presents the posterior distributions for the
neutron-decay scenario. Most nuclear-matter parameters are reasonably well
localized, whereas $E_0$ remains broadly distributed and $\rho_0$ is
weighted toward the upper part of its prior range. The most notable
model-dependent result is the skewness coefficient,
$Q_0=-644.18^{+112.36}_{-139.28}\,\mathrm{MeV}$, which is shifted toward
more negative values compared with the Higgs-portal model. The symmetry
energy parameters are centered at
$J_0=33.25^{+1.40}_{-1.52}\,\mathrm{MeV}$ and
$L_0=58.61^{+7.34}_{-6.25}\,\mathrm{MeV}$ and exhibit a clear positive
correlation. The neutron-decay model also favors a somewhat larger
effective-mass ratio, although the corresponding credible region overlaps
with that of the Higgs-portal model. All quoted uncertainties represent
68\% credible intervals.

In contrast to the nuclear-matter parameters, the dark-sector
self-interaction remains weakly constrained. We obtain
$\log_{10}(G_\chi/\mathrm{fm}^{2})=2.73^{+0.20}_{-0.41}$, corresponding
to a median $G_\chi\simeq537\,\mathrm{fm}^{2}$ and a 68\% credible range
of approximately $209$--$851\,\mathrm{fm}^{2}$. Since this interval
covers a substantial portion of the adopted prior and the prior is
uniform in $G_\chi$ rather than in $\log G_\chi$, the concentration at
large logarithmic values should not be interpreted as a precise
measurement of the interaction strength. The pairwise posteriors additionally reveal an anticorrelation between
$Q_0$ and $M^*/M$, but no strong correlation between $G_\chi$ and the
nuclear-matter parameters. Overall, the neutron-decay mechanism modifies
the preferred hadronic parameter region, particularly through a more
negative $Q_0$, while the strength of the dark-sector interaction itself
remains poorly determined by the present data.

\begin{figure*}
    \centering
    \includegraphics[width=\linewidth]{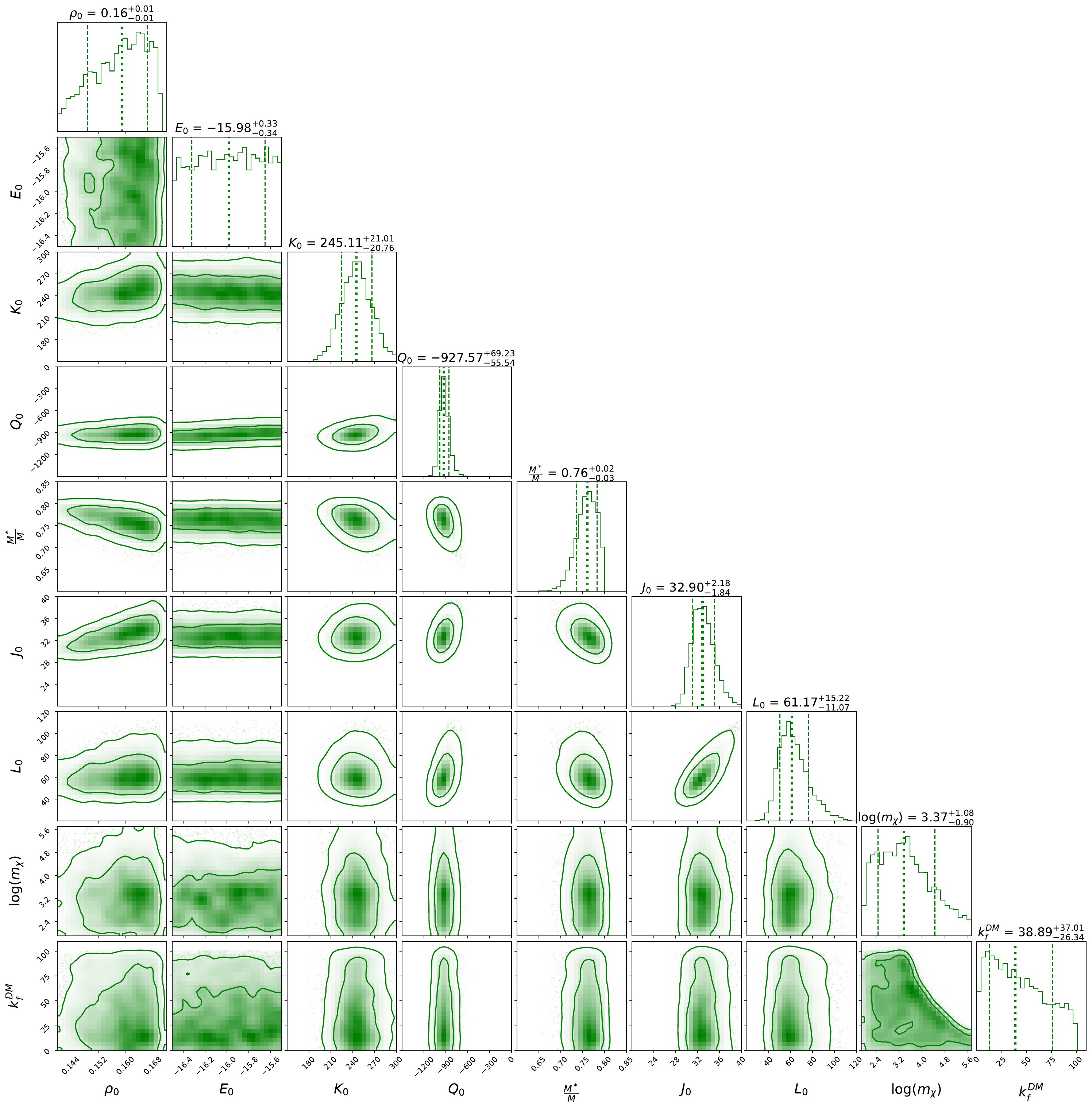}
    \caption{Marginalized posterior distributions for the ALP-mediated model.
The plotting conventions are the same as in
Fig.~\ref{fig:corner_Higgs}. The inferred dark-sector parameters are
$\log_{10}(m_\chi/{\rm MeV})$ and $k_F^{\rm DM}$ in MeV.}
    \label{fig:corner_ALP}
\end{figure*}

Figure~\ref{fig:corner_ALP} presents the posterior distributions for the
ALP-mediated model. Among the nuclear-matter parameters, the most
distinctive result is the comparatively narrow posterior of the skewness
coefficient, $Q_0=-927.57^{+69.23}_{-55.54}\,\mathrm{MeV}$, which favors
substantially more negative values than in the other two models. The
incompressibility and effective-mass ratio are also reasonably well
localized around $K_0=245.11^{+21.01}_{-20.76}\,\mathrm{MeV}$ and
$M^*/M=0.76^{+0.02}_{-0.03}$, respectively. In contrast, $E_0$ remains
broadly distributed over its prior range. The symmetry-energy parameters
are inferred to be $J_0=32.90^{+2.18}_{-1.84}\,\mathrm{MeV}$ and
$L_0=61.17^{+15.22}_{-11.07}\,\mathrm{MeV}$, although the posterior of
$L_0$ is wider and more asymmetric. The pairwise distributions show the
expected positive correlation between $J_0$ and $L_0$, along with an
anticorrelation between $Q_0$ and $M^*/M$. All quoted uncertainties
represent 68\% credible intervals. The dark-sector parameters remain considerably less constrained. The
posterior gives
$\log_{10}(m_\chi/\mathrm{MeV})=3.37^{+1.08}_{-0.90}$ and
$k_F^{\mathrm{DM}}=38.89^{+37.01}_{-26.34}\,\mathrm{MeV}$. The former
corresponds to a median dark-matter mass of approximately
$2.34\,\mathrm{GeV}$, but its 68\% credible interval extends from about
$0.30$ to $28.2\,\mathrm{GeV}$. A broad anticorrelation is visible between
$m_\chi$ and $k_F^{\mathrm{DM}}$, with large Fermi momenta predominantly
allowed for smaller dark-matter masses. This behavior is physically
consistent with $n_\chi\propto(k_F^{\mathrm{DM}})^3$: an increase in
$m_\chi$ must generally be accompanied by a reduction in the dark-matter
number density to limit its total contribution to the stellar EOS.

Considering all three scenarios, the common constraints produce broadly
overlapping symmetry-energy posteriors, while $Q_0$ becomes progressively
more negative from the Higgs-portal model to the neutron-decay and
ALP-mediated models. This model dependence indicates that the effects of
different dark sectors are partly absorbed through changes in the
high-density hadronic EOS. Meanwhile, the broad dark-sector posteriors in
all three cases show that present observations constrain the bulk stellar
response more effectively than the underlying microscopic interaction.

\begin{figure}
    \centering
    \includegraphics[width=0.5\linewidth]{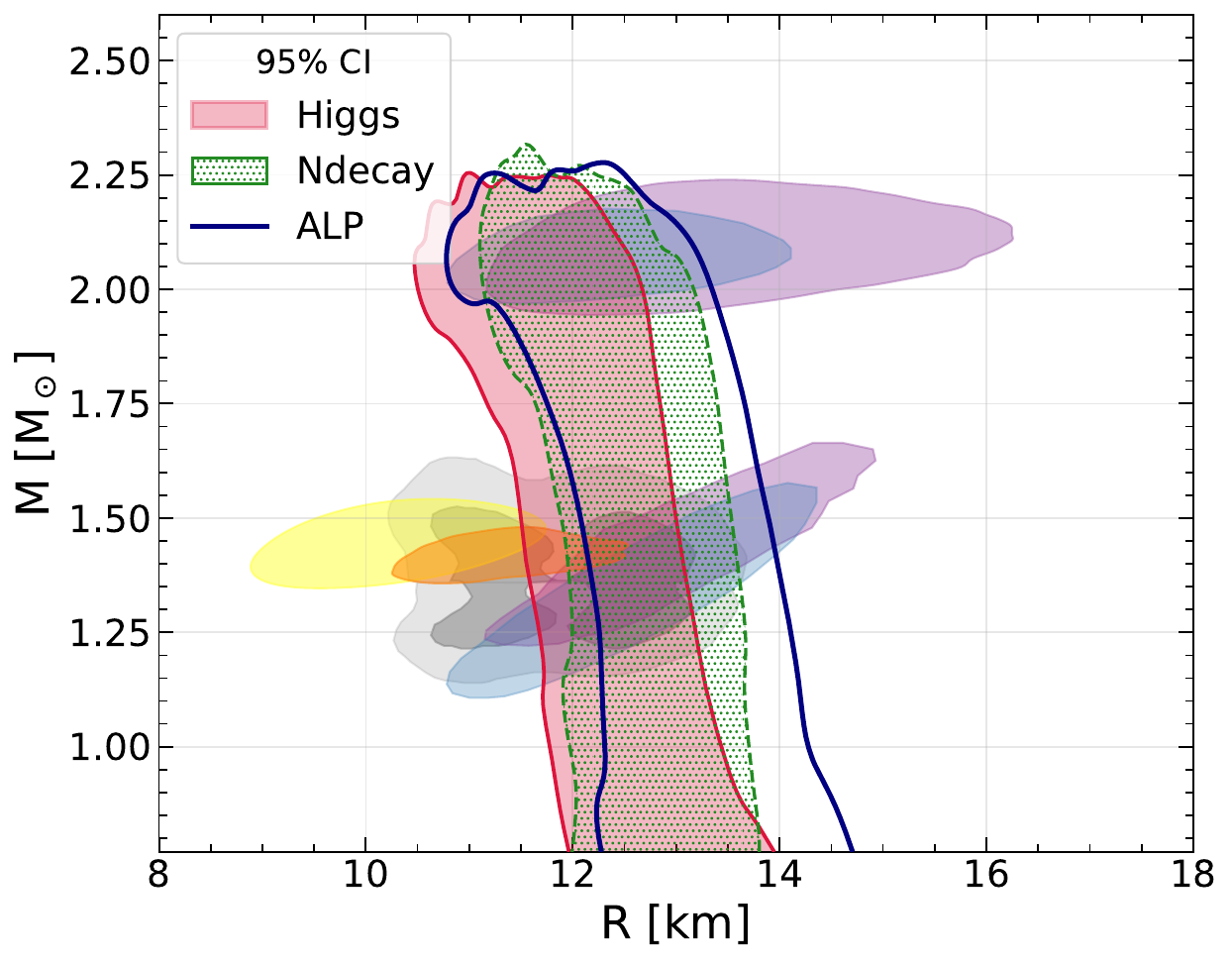}\includegraphics[width=0.5\linewidth]{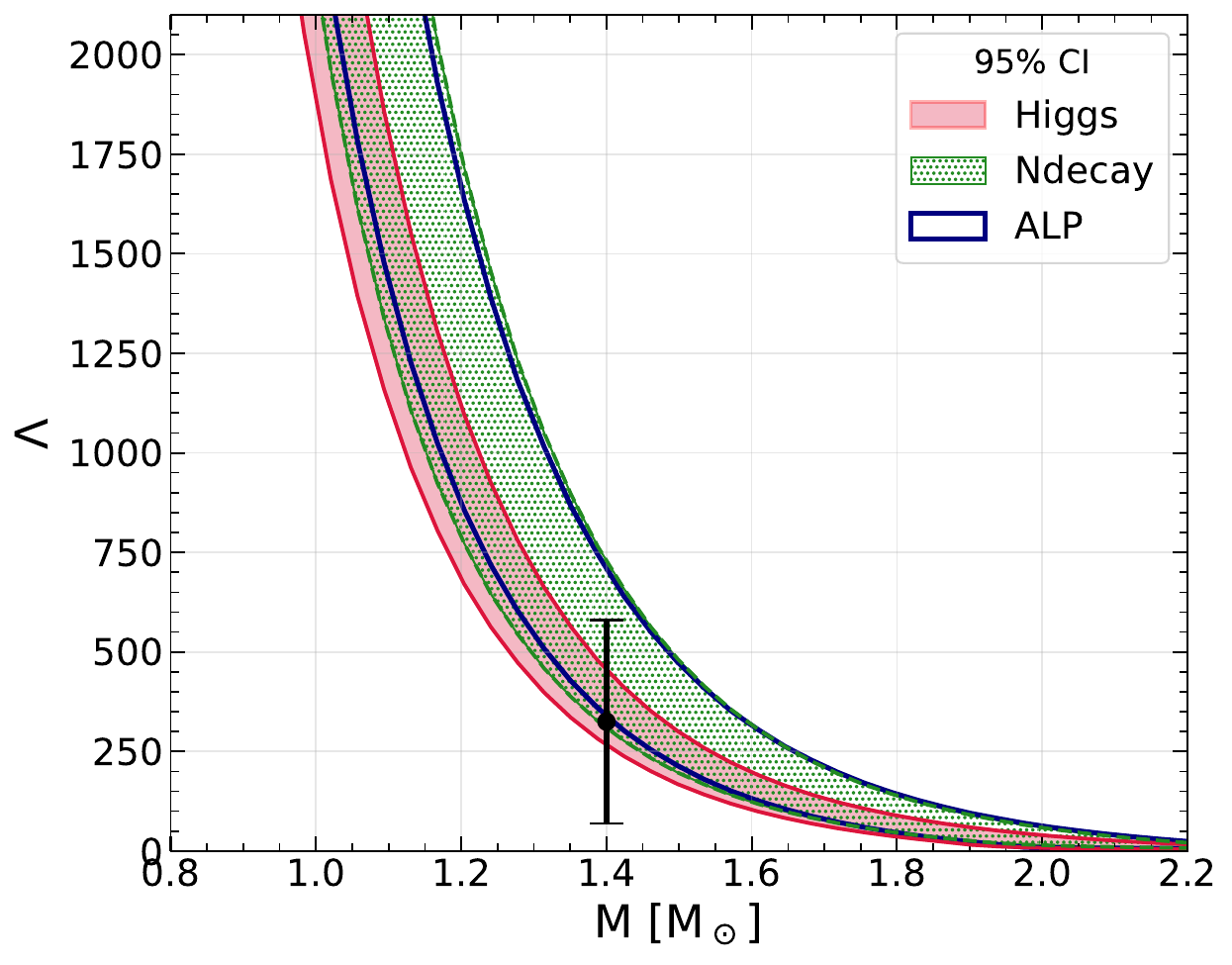}
    \caption{Posterior predictions for the mass--radius relation (left) and
mass--tidal-deformability relation (right). The 95\% credible regions
correspond to the Higgs-portal (red), neutron-decay (green dotted), and
ALP-mediated (blue) models. The left panel also shows the NICER
mass--radius posteriors for PSR J0740+6620, PSR J0030+0451,
PSR J0437-4715, and PSR J0614-3329, together with the GW170817
constraints shown in grey. In the right panel, the black marker and error
bar denote the adopted GW170817 constraint at $1.4\,M_\odot$.}
    \label{fig:MRplot}
\end{figure}

Figure~\ref{fig:MRplot} shows the macroscopic stellar properties obtained
by propagating the posterior EOS ensembles through the TOV equations.
Despite their different microscopic descriptions, the three models predict
very similar maximum masses: Table~\ref{tab:NSprop} gives median values of
$2.09$, $2.13$, and $2.12\,M_\odot$ for the Higgs-portal, neutron-decay,
and ALP-mediated models, respectively. This similarity indicates that the
high-density pressure is strongly regulated by the massive-pulsar
constraints, with changes induced by the dark sector partly accommodated
through the inferred hadronic EOS. More visible differences occur around
the canonical mass, where the median radii are $12.24$, $12.39$, and
$12.87$ km, respectively. The smaller radius predicted by the Higgs-portal
model leads to a lower median tidal deformability,
$\Lambda_{1.4}\simeq361$, whereas the neutron-decay and ALP-mediated models
give comparable values of approximately $500$ and $495$. This behavior
follows from the strong compactness dependence
$\Lambda=(2/3)k_2(R/M)^5$, although differences in the Love number and
internal density profile prevent the ordering in radius from being
transferred exactly to $\Lambda$. The 95\% credible regions nevertheless
overlap substantially and intersect the adopted GW170817 constraint, while
the smallest-radius NICER regions are only partially covered. Since these
observations enter the inference, the overlap represents posterior
consistency rather than an independent test. Overall, present mass--radius
and tidal-deformability measurements constrain the bulk EOS response but
do not distinguish the three dark-sector mechanisms.

\begin{table}[]
    \centering
    \renewcommand{\arraystretch}{1.5}
    \begin{tabular}{|c|c|c|c|c|c|c|}
        \hline
        Model &  M$_{\text{max}}(M_\odot)$ & R$_{\text{max}}$(km) & R$_{1.4}$(km) & $\Lambda_{1.4}$  & $\nu_{f,1.4}$(kHz)  & $\tau_{f,1.4}$(s) \\
        \hline
        Higgs & $2.09^{+0.26}_{-0.18}$ & $10.78^{+0.65}_{-0.62}$ & $12.24^{+0.49}_{-0.79}$  & $361.46^{+93.66}_{-95.36}$  & $1.83^{+0.13}_{-0.08}$ & $0.21^{+0.02}_{-0.02}$\\
         \hline
        N-decay & $2.13^{+0.2}_{-0.17}$  & $11.22^{+0.35}_{-0.44}$  & $12.39^{+0.65}_{-0.82}$  & $500.00^{+226.03}_{-191.88}$  & $1.71^{+0.19}_{-0.13}$ & $0.24^{+0.09}_{-0.04}$\\
         \hline
        ALP & $2.12^{+0.21}_{-0.16}$ & $11.09^{+0.72}_{-0.66}$ & $12.87^{+0.78}_{-0.77}$ & $495.03^{+214.68}_{-156.91}$ & $1.71^{+0.15}_{-0.12}$ & $0.24^{+0.04}_{-0.03}$\\
        \hline
    \end{tabular}
    \caption{NS properties for the three DM models: Median values of maximum mass M$_\text{max}$, radius corresponding to M$_\text{max}$, radius, tidal deformability, f-mode frequency and damping time at star mass M=1.4M$_\odot$. }
    \label{tab:NSprop}
\end{table}

\begin{figure}
    \centering
    \includegraphics[width=0.5\linewidth]{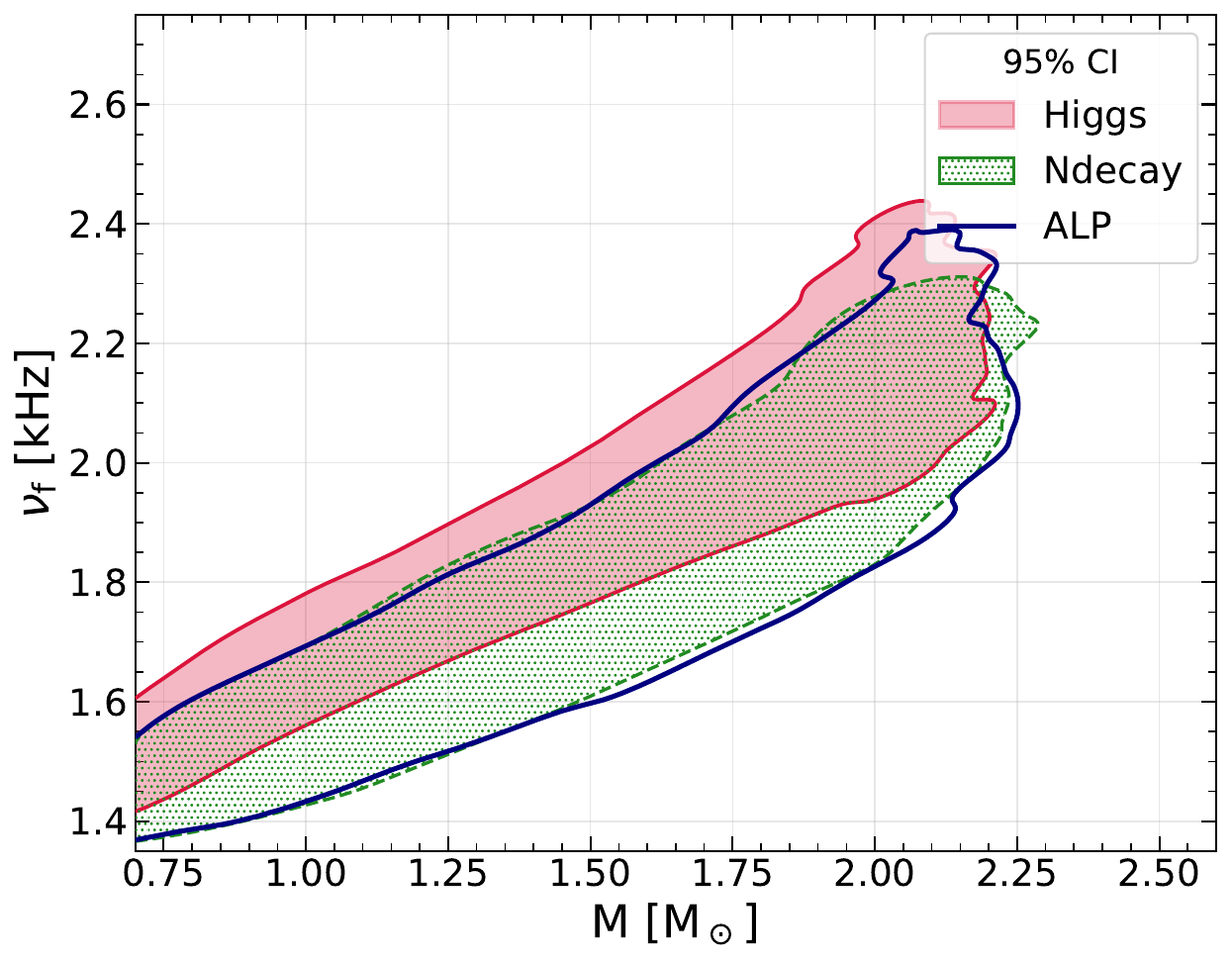}\includegraphics[width=0.5\linewidth]{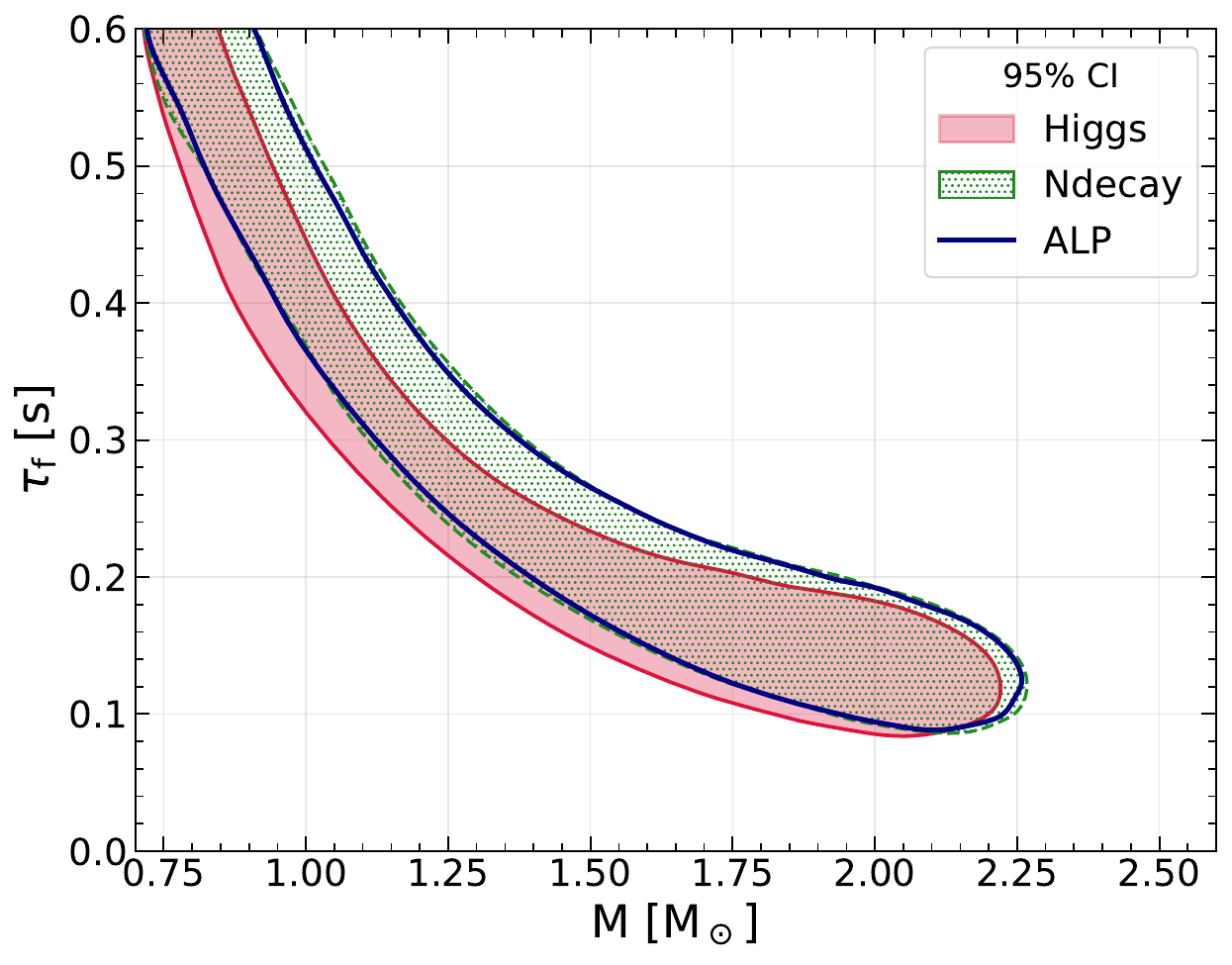}
    \caption{95\% CI for the f-mode frequency $\nu_f$ (left panel) and damping time $\tau_f$ (right panel) for the 3 DM models.}
    \label{fig:f_tau}
\end{figure}

Figure~\ref{fig:f_tau} shows that the $f$-mode frequency generally
increases with stellar mass, while the damping time decreases as the star
becomes more compact and gravitational-wave emission becomes more
efficient. At $1.4\,M_\odot$, Table~\ref{tab:NSprop} gives
$\nu_{f,1.4}=1.83$ kHz and $\tau_{f,1.4}=0.21$ s for the Higgs-portal
model, compared with approximately $1.71$ kHz and $0.24$ s for both the
neutron-decay and ALP-mediated models. This ordering follows directly from
the mass--radius results: the smaller radius predicted by the Higgs-portal
model produces a larger mean density, and hence a higher $f$-mode
frequency and shorter damping time, whereas the larger radii of the other
two models lead to lower frequencies and slower damping. The 95\%
credible regions still overlap, particularly for the neutron-decay and
ALP-mediated scenarios, showing that their different microscopic
interactions generate nearly degenerate oscillation properties. Thus,
$f$-mode measurements provide a dynamical probe complementary to the
static observables, but a sufficiently precise measurement would be
required to distinguish the Higgs-portal model, while separating the
neutron-decay and ALP-mediated scenarios would remain more challenging.

\subsection{Bayesian model comparison}

\begin{table}[h]
    \centering
    \begin{tabular}{|c|c||c|c|}
        \hline
        Model & log evidence & Model & log evidence\\
        \hline
        Higgs portal & -62.48 & FDM & -62.84\\
        N-decay & -59.22 & BDM1 & -62.41\\
        ALP mediator & -60.02 & BDM2 & -62.39\\
        \hline
    \end{tabular}
    \caption{Log evidence values for the DM models}
    \label{tab:lnZ}
\end{table}

\begin{figure}
    \centering
    \includegraphics[width=0.8\linewidth]{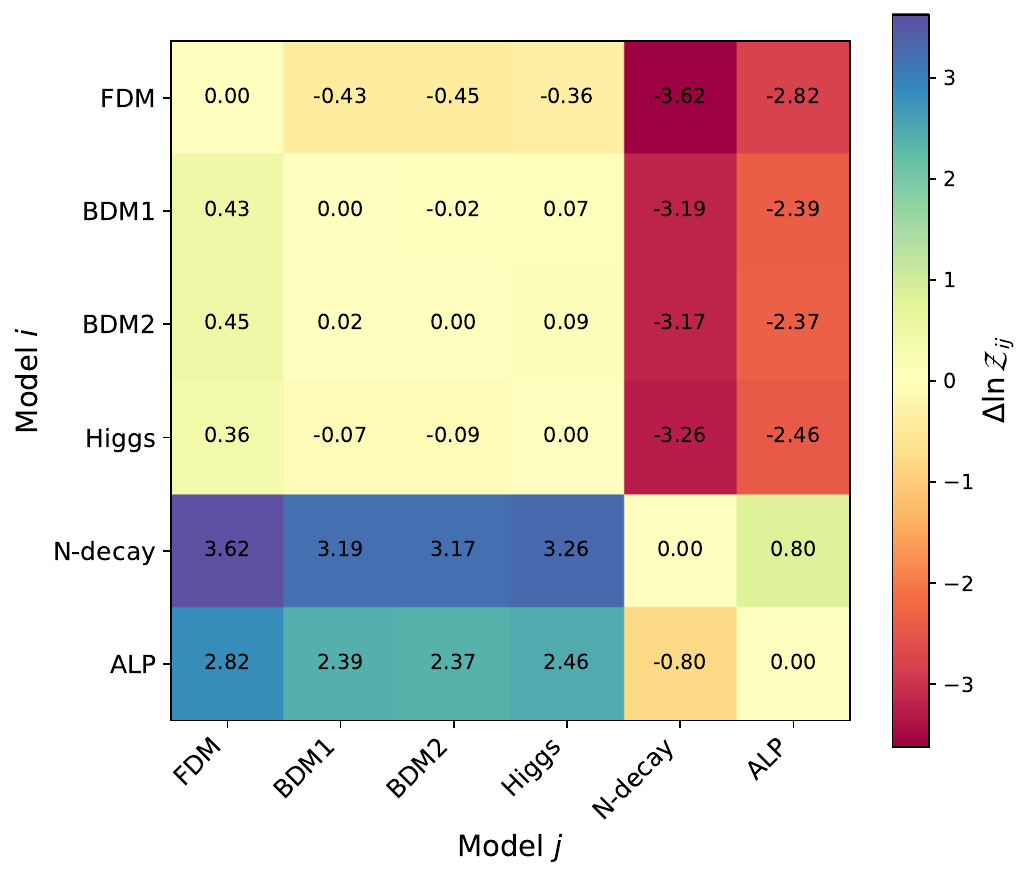}
    \caption{Matrix showing \(\Delta\ln\mathcal Z_{ij}\) where model along the column is \(i\) and along the row is \( j\).}
    \label{fig:bayes_heatmap}
\end{figure}

In our previous work \cite{Arvikar:2026}, we performed a Bayesian analysis to compare three two-fluid models of DM admixed NS incorporating fermionic and bosonic DM: a fermionic DM model with repulsive vector interactions (FDM), an isotropic bosonic DM model (BDM1), and an anisotropic bosonic DM model (BDM2), as described in Ref. \cite{Arvikar:2026}. We found that the current astrophysical observations do not significantly distinguish between these models. In the present study, we extend this analysis to investigate whether the available data can discriminate between DM models that interact with ordinary matter solely through gravity and those that also involve additional potential-mediated interactions. We compare these models through Bayes' factors, as described below.

The two-fluid FDM, BDM1, and BDM2 models of Ref.~\cite{Arvikar:2026} were reanalysed using the same seven hadronic parameters and nuclear and astrophysical likelihoods adopted for the single-fluid scenarios. For two-fluid configurations, the visible hadronic radius \(R_H\) enters the NICER likelihood, while the tidal response is calculated from the complete stellar configuration with outer radius \(R_T\). Each evidence is obtained by marginalizing the likelihood over the corresponding normalized parameter prior, allowing comparisons between models with different parameter dimensions and fluid descriptions. Figure~\ref{fig:bayes_heatmap} presents the pairwise log Bayes factors, \(\ln B_{ij}=\ln\mathcal Z_i-\ln\mathcal Z_j\), calculated from the evidences in Table~\ref{tab:lnZ}; positive entries favor the row model over the column model.

The heatmap shows that the neutron-decay and ALP-mediated models have higher evidences than the other four scenarios, although their mutual separation is small. Neutron decay has the highest evidence, but its advantage over ALP is only \(\ln B_{\rm N,ALP}=0.80\), corresponding to \(B_{\rm N,ALP}\simeq2.2\), and therefore provides no clear discrimination between these two mechanisms. Relative to the Higgs-portal and two-fluid models, neutron decay yields Bayes factors of approximately \(24\)--\(37\), while ALP yields factors of approximately \(11\)--\(17\). The Higgs-portal, FDM, BDM1, and BDM2 models have closely spaced evidences, with pairwise differences satisfying \(|\ln B_{ij}|\leq0.45\), indicating little preference among them under the adopted priors. In particular, the proximity of the single-fluid Higgs-portal model to all three two-fluid models shows that the evidence ranking does not follow a simple division between fluid descriptions. These comparisons reflect the combined effects of microscopic interactions, abundance prescriptions, and parameter priors; the evidence values alone do not identify which of these features drives the ranking. A comparison with a consistently analysed no-DM model would additionally be required to determine whether the data favor a dark component itself.

\section{Conclusion}\label{conclusion}

We have investigated Higgs-portal, neutron-decay, and ALP-mediated dark-matter scenarios within a common Bayesian framework combining nuclear constraints, chiral effective field theory, NICER measurements, and GW170817. The analysis constrains several nuclear-matter parameters, while the individual dark-sector parameters remain weakly determined. In particular, the broad, correlated posteriors of $m_\chi$ and $k_F^{\rm DM}$ in the Higgs-portal and ALP scenarios indicate that present observations constrain their combined contribution to stellar structure more effectively than either parameter separately. The model-dependent shifts in the hadronic parameters, especially the substantially more negative $Q_0$ favored by the ALP scenario, also demonstrate that the inferred hadronic equation of state is sensitive to the assumed dark-sector physics. Separating these contributions therefore remains a central limitation of the present inference.

The three scenarios yield substantially overlapping mass--radius and tidal-deformability predictions despite their different microscopic descriptions. For canonical-mass stars, the Higgs-portal model tends to predict greater compactness, higher $f$-mode frequencies, and shorter damping times, whereas the neutron-decay and ALP predictions are more closely grouped. These oscillation properties provide prospective observational tests, rather than additional constraints used in the present analysis. The Bayesian comparison with consistently reanalysed two-fluid models nevertheless reveals differences in their overall support: under the adopted priors, the neutron-decay and ALP scenarios have higher evidences than the Higgs-portal and three two-fluid alternatives, although the preference for neutron decay over ALP is weak. This ranking concerns the complete models, including their interactions, abundance prescriptions, and parameter priors; it does not establish a general preference for the single-fluid formalism.

Further progress requires reducing the freedom of the hadronic EOS to accommodate different dark-sector contributions. Improved nuclear constraints, together with more precise radius and tidal-deformability measurements across a wider stellar-mass range, could help resolve this ambiguity. Future measurements of $f$-mode frequencies and damping times would offer complementary dynamical information, although the overlap of the present predictions makes joint inference with other observables essential. Systematic tests of prior sensitivity and a comparison with a consistently analysed no-DM model are also needed to determine whether any preference among dark-matter scenarios translates into evidence for a dark component itself.

\section*{Acknowledgments}
Authors acknowledge the computing facilities at BITS Pilani-Hyderabad Campus, used for the analyses. PA is thankful to Debanjan Guha Roy for the fruitful discussions. P.~Thakur is supported by the National Research Foundation of Korea (NRF) grant funded by the Korea government (MSIT) (No.~RS-2024-00457037). His work was supported (in part) by the Yonsei University Research Fund(Yonsei University Frontier Fellowship for Postdoctoral Researchers) of 2025.

\newpage
\bibliographystyle{JHEP} 
\bibliography{DM_NS}

@article{Kouvaris_2010,
   title={Can neutron stars constrain dark matter?},
   volume={82},
   ISSN={1550-2368},
   url={http://dx.doi.org/10.1103/PhysRevD.82.063531},
   number={6},
   pages = {063531},
   journal={Phys. Rev. D},
   publisher={American Physical Society (APS)},
   author={Kouvaris, Chris and Tinyakov, Peter},
   year={2010},
   month=sep }

@article{das2020dark,
    author = {Das, Arpan and Malik, Tuhin and Nayak, Alekha C.},
    title = {Dark matter admixed neutron star properties in light of gravitational wave observations: A two fluid approach},
    journal = {Phys. Rev. D},
    volume = {105},
    number = {12},
    pages = {123034},
    year = {2022}
}

@article{Das_2019,
   title="{Confronting nuclear equation of state in the presence of dark matter using GW170817 observation in relativistic mean field theory approach}",
   volume={99},
   pages = {043016},
   ISSN={2470-0029},   
   number={4},
   journal={Phys. Rev. D},
   publisher={American Physical Society (APS)},
   author={Das, Arpan and Malik, Tuhin and Nayak, Alekha C.},
   year={2019},
   month=feb }

@article{Dutra_2014,
   title={Relativistic mean-field hadronic models under nuclear matter constraints},
   volume={90},
   ISSN={1089-490X},
   number={5},
   journal={Phy. Rev. C},
   pages = {055203},
   publisher={American Physical Society (APS)},
   author={Dutra {\sl et al.}, M.},
   year={2014},
   month=nov }

@article{Xiang_2014,
   title={Effects of fermionic dark matter on properties of neutron stars},
   volume={89},
   ISSN={1089-490X},
   number={2},
   journal={Phys. Rev. C},
   publisher={American Physical Society (APS)},
   author={Xiang, Qian Fei and Jiang, Wei Zhou and Zhang, Dong Rui and Yang, Rong Yao},
   year={2014},
   month=feb }

@article{thakur2023exploring,
  title = {Exploring robust correlations between fermionic dark matter model parameters and neutron star properties: A two-fluid perspective},
  author = {Thakur, Prashant and Malik, Tuhin and Das, Arpan and Jha, T. K. and Provid\^encia, Constan\ifmmode \mbox{\c{c}}\else \c{c}\fi{}a},
  journal = {Phys. Rev. D},
  volume = {109},
  issue = {4},
  pages = {043030},
  numpages = {15},
  year = {2024},
  month = {Feb},
  publisher = {American Physical Society},
  doi = {10.1103/PhysRevD.109.043030},
  url = {https://link.aps.org/doi/10.1103/PhysRevD.109.043030}
}

@article{Routaray_2023,
   title="{Investigating dark matter-admixed neutron stars with {NITR} equation of state in light of PSR J0952-0607}",
   volume={2023},
   ISSN={1475-7516},
   number={10},
   journal={JCAP},
   publisher={IOP Publishing},
   author={Routaray, Pinku and Mohanty, Sailesh Ranjan and Das, H.C. and Ghosh, Sayantan and Kalita, P.J. and Parmar, Vishal and Kumar, Bharat},
   year={"2023"},
   month=oct, pages={073} }

@article{Li_2012,
   title={Too massive neutron stars: The role of dark matter?},
   volume={37},
   ISSN={0927-6505},
   url={http://dx.doi.org/10.1016/j.astropartphys.2012.07.006},
   DOI={10.1016/j.astropartphys.2012.07.006},
   journal={Astropart. Phys.},
   publisher={Elsevier BV},
   author={Li, Ang and Huang, Feng and Xu, Ren Xin},
   year={2012},
   month=sep, pages={70} }

@article{Venneti_2024,
   title={Unraveling the global behavior of equation of state by explicit finite nuclei constraints},
   volume={854},
   ISSN={0370-2693},
   url={http://dx.doi.org/10.1016/j.physletb.2024.138756},
   DOI={10.1016/j.physletb.2024.138756},
   journal={Phys. Lett. B},
   publisher={Elsevier BV},
   author={Venneti, Anagh and Gautam, Sakshi and Banik, Sarmistha and Agrawal, B.K.},
   year={2024},
   month=jul, pages={138756} }

@article{Choudhury:2024xbk,
    author = "Choudhury, Devarshi and others",
    title = "{A NICER View of the Nearest and Brightest Millisecond Pulsar: PSR J0437\textendash{}4715}",
    journal = "Astrophys. J. Lett.",
    volume = "971",
    number = "1",
    pages = "L20",
    year = "2024"
}

@article{Abbott_2018,
  title = "{GW170817: Measurements of Neutron Star Radii and Equation of State}",
  author = {Abbott {\sl et al.}, B. P. },
  collaboration = {The LIGO Scientific and the Virgo Collaboration},
  journal = {Phys. Rev. Lett.},
  volume = {121},
  issue = {16},
  pages = {161101},
  numpages = {16},
  year = {2018},
  month = {Oct},
  publisher = {APS},
  doi = {10.1103/PhysRevLett.121.161101},
  url = {https://link.aps.org/doi/10.1103/PhysRevLett.121.161101}
}

@article{Miao_2022,
   title={Dark Matter Admixed Neutron Star Properties in the Light of X-Ray Pulse Profile Observations},
   volume={936},
   ISSN={1538-4357},
   url={http://dx.doi.org/10.3847/1538-4357/ac8544},
   DOI={10.3847/1538-4357/ac8544},
   number={1},
   journal={Astrophys. J.},
   publisher={American Astronomical Society},
   author={Miao, Zhiqiang and Zhu, Yaofeng and Li, Ang and Huang, Feng},
   year={2022},
   month=aug, pages={69} }

@article{Shirke23,
    author = {Shirke, Swarnim and Ghosh, Suprovo and Chatterjee, Debarati and Sagunski, Laura and Schaffner-Bielich, J\"urgen},
    title = "{R-modes as a new probe of dark matter in neutron stars}",
    journal = "JCAP",
    volume = "12",
    pages = "008",
    year = "2023"
}

@article{Karkevandi2021,
    author = "Karkevandi, Davood Rafiei and Shakeri, Soroush and Sagun, Violetta and Ivanytskyi, Oleksii",
    title = "{Bosonic dark matter in neutron stars and its effect on gravitational wave signal}",
    journal = "Phys. Rev. D",
    volume = "105",
    number = "2",
    pages = "023001",
    year = "2022"
}

@article{AnkitKumar_2024,
    author = "Kumar, Ankit and Sotani, Hajime",
    title = "{Constraints on the parameter space in dark matter admixed neutron stars}",
    journal = "Phys. Rev. D",
    volume = "110",
    number = "6",
    pages = "063001",
    year = "2024"
}

@article{Miller_2019,
   title="{PSR J0030+0451 Mass and Radius from NICER Data and Implications for the Properties of Neutron Star Matter}",
   volume={887},
   ISSN={2041-8213},
   url={http://dx.doi.org/10.3847/2041-8213/ab50c5},
   DOI={10.3847/2041-8213/ab50c5},
   number={1},
   journal={Astrophys. J. Lett.},
   publisher={American Astronomical Society},
   author={Miller {\sl et al.}, M. C.},
   year={2019},
   month=dec, pages={L24} }

@article{Riley_2019,
   title="{A NICER view of PSR J0030+0451: Millisecond Pulsar Parameter Estimation}",
   volume={887},
   ISSN={2041-8213},
   url={http://dx.doi.org/10.3847/2041-8213/ab481c},
   DOI={10.3847/2041-8213/ab481c},
   number={1},
   journal={Astrophys. J. Lett.},
   publisher={American Astronomical Society},
   author={Riley {\sl et al.}, T. E.}, 
   year={2019},
   month=dec, pages={L21} }

@article{Riley_2021,
   title="{A NICER View of the Massive Pulsar PSR J0740+6620 Informed by Radio Timing and XMM-Newton Spectroscopy}",
   volume={918},
   ISSN={2041-8213},
   url={http://dx.doi.org/10.3847/2041-8213/ac0a81},
   DOI={10.3847/2041-8213/ac0a81},
   number={2},
   journal={Astrophys. J. Lett.},
   publisher={American Astronomical Society},
   author={Riley {\sl et al.}, Thomas E.},
   year={2021},
   month=sep, pages={L27} }

@article{Miller_2021,
   title="{The Radius of PSR J0740+6620 from NICER and XMM-Newton Data}",
   volume={918},
   ISSN={2041-8213},
   url={http://dx.doi.org/10.3847/2041-8213/ac089b},
   DOI={10.3847/2041-8213/ac089b},
   number={2},
   journal={Astrophys. J. Lett.},
   publisher={American Astronomical Society},
   author={Miller {\sl et al.}, M. C.}, 
   year={2021},
   month=sep, pages={L28} }

@article{Rutherford_2023,
   title={Constraining bosonic asymmetric dark matter with neutron star mass-radius measurements},
   volume={107},
   ISSN={2470-0029},
   number={10},
   journal={Phys. Rev. D},
   pages = {103051},
   publisher={American Physical Society (APS)},
   author={Rutherford, Nathan and Raaijmakers, Geert and Prescod-Weinstein, Chanda and Watts, Anna},
   year={2023},
   month=may }

@article{Husain_2021,
author = {Husain, Wasif and Thomas, Anthony W.},
title = {Possible nature of dark matter},
journal = {JCAP},
volume = {2021},
pages = {086},
year = {2021},
month = {oct},
publisher = {IOP Publishing}
}

@ARTICLE{Aprile_2012,
       author = {{Aprile} {\sl et al.}, E.}, 
          title = "{The XENON100 dark matter experiment}",
          collaboration = "{Xenon100 Collaboration}", 
          journal = {Astroparticle Physics},
             year = 2012,
            month = apr,
           volume = {35},
           number = {9},
            pages = {573-590}
}

@article{Wang_2020,
    author = "Wang, Qiuhong and others",
    collaboration = "PandaX-II",
    title = "{Results of dark matter search using the full PandaX-II exposure}",
    eprint = "2007.15469",
    archivePrefix = "arXiv",
    primaryClass = "astro-ph.CO",
    doi = "10.1088/1674-1137/abb658",
    journal = "Chin. Phys. C",
    volume = "44",
    number = "12",
    pages = "125001",
    year = "2020"
}

@article{DONATO2014,
title = {Indirect searches for dark matter},
journal = {Phys. Dark Universe},
volume = {4},
pages = {41},
year = {2014},
author = {Fiorenza Donato}
}

@article{Heros2020,
    author = "P\'erez de los Heros, Carlos",
    title = "{Status, Challenges and Directions in Indirect Dark Matter Searches}",
    journal = "Symmetry",
    volume = "12",
    number = "10",
    pages = "1648",
    year = "2020"
}

@article{Biondini:2023,
    author = "Biondini, Simone and Bollig, Julian and Vogl, Stefan",
    title = "{Indirect detection of dark matter with (pseudo)-scalar interactions}",
    journal = "JHEP",
    year = "2024",
    volume = "04",
    pages = "050"
}

@book{stuart1994kendalls,
  author    = {Stuart, A. and Ord, J.K.},
  title     = {"Kendall’s Advanced Theory of Statistics. Volume 1. Distribution Theory"},
  edition   = {Sixth},
  year      = {1994},
  publisher = {Edward Arnold},
  address   = {London}
}

@article{Ellis_2018,
  title = {Dark matter effects on neutron star properties},
  author = {Ellis, John and H\"utsi, Gert and Kannike, Kristjan and Marzola, Luca and Raidal, Martti and Vaskonen, Ville},
  journal = {Phys. Rev. D},
  volume = {97},
  issue = {12},
  pages = {123007},
  numpages = {7},
  year = {2018},
  month = {Jun}
}

@article{Shawqi:2024,
    author = "Shawqi, Shafayat and Morsink, Sharon M.",
    title = "{Interpreting Mass and Radius Measurements of Neutron Stars with Dark Matter Halos}",
    eprint = "2406.03332",
    archivePrefix = "arXiv",
    primaryClass = "astro-ph.HE",
    reportNumber = "INT-PUB-24-025",
    doi = "10.3847/1538-4357/ad77c1",
    journal = "Astrophys. J.",
    volume = "975",
    number = "1",
    pages = "123",
    year = "2024"
}

@article{Klangburam:2025,
    author = "Klangburam, Tanech and Pongkitivanichkul, Chakrit",
    title = "{Axion-Like Particle Mediated Dark Matter and Neutron Star Properties in the QHD Model}",
    eprint = "2503.12430",
    archivePrefix = "arXiv",
    primaryClass = "astro-ph.HE",
    month = "3",
    year = "2025"
}

@article{Fornal:2020gto,
    author = "Fornal, Bartosz and Grinstein, Benjam\'\i{}n",
    title = "{Neutron\textquoteright{}s dark secret}",
    eprint = "2007.13931",
    archivePrefix = "arXiv",
    primaryClass = "hep-ph",
    doi = "10.1142/S0217732320300190",
    journal = "Mod. Phys. Lett. A",
    volume = "35",
    number = "31",
    pages = "2030019",
    year = "2020"
}

@article{Arvikar:2025,
    author = "Arvikar, Payaswinee and Gautam, Sakshi and Venneti, Anagh and Banik, Sarmistha",
    title = "{Exploring fermionic dark matter admixed neutron stars in the light of astrophysical observations}",
    eprint = "2506.20736",
    archivePrefix = "arXiv",
    primaryClass = "astro-ph.HE",
    doi = "10.1103/kwwx-54wq",
    journal = "Phys. Rev. D",
    volume = "112",
    number = "2",
    pages = "023021",
    year = "2025"
}

@article{Arvikar:2026,
    author = "Arvikar, Payaswinee and Gautam, Sakshi and Venneti, Anagh and Banik, Sarmistha",
    title = "{Fermionic versus Bosonic Dark Matter in Neutron Stars: A bayesian study with multi-density constraints}",
    eprint = "2512.13574",
    archivePrefix = "arXiv",
    primaryClass = "astro-ph.CO",
    doi = "10.1088/1475-7516/2026/03/012",
    journal = "JCAP",
    volume = "03",
    pages = "012",
    year = "2026"
}

@article{PThakur_2026,
  title = {Constraining axionlike particle mediated dark matter with observational constraints: A statistical and machine learning approach},
  author = {Thakur, Prashant and Taridalu, Aravind and Rather, Ishfaq Ahmad and Klangburam, Tanech and Pongkitivanichkul, Chakrit},
  journal = {Phys. Rev. D},
  volume = {113},
  issue = {2},
  pages = {023001},
  numpages = {22},
  year = {2026},
  month = {Jan},
  publisher = {American Physical Society},
  doi = {10.1103/j3mm-zjsv},
  url = {https://link.aps.org/doi/10.1103/j3mm-zjsv}
}

@article{Pattie:2017vsj,
    author = "Pattie, Jr., R. W. and others",
    title = "{Measurement of the neutron lifetime using a magneto-gravitational trap and in situ detection}",
    eprint = "1707.01817",
    archivePrefix = "arXiv",
    primaryClass = "nucl-ex",
    doi = "10.1126/science.aan8895",
    journal = "Science",
    volume = "360",
    number = "6389",
    pages = "627--632",
    year = "2018"
}

@article{Yue:2013qrc,
    author = "Yue, A. T. and Dewey, M. S. and Gilliam, D. M. and Greene, G. L. and Laptev, A. B. and Nico, J. S. and Snow, W. M. and Wietfeldt, F. E.",
    title = "{Improved Determination of the Neutron Lifetime}",
    eprint = "1309.2623",
    archivePrefix = "arXiv",
    primaryClass = "nucl-ex",
    doi = "10.1103/PhysRevLett.111.222501",
    journal = "Phys. Rev. Lett.",
    volume = "111",
    number = "22",
    pages = "222501",
    year = "2013"
}

@article{Shirke:2024ymc,
    author = {Shirke, Swarnim and Pradhan, Bikram Keshari and Chatterjee, Debarati and Sagunski, Laura and Schaffner-Bielich, J\"urgen},
    title = "{Effects of dark matter on f-mode oscillations of neutron stars}",
    eprint = "2403.18740",
    archivePrefix = "arXiv",
    primaryClass = "gr-qc",
    reportNumber = "LIGO-P2400135",
    doi = "10.1103/PhysRevD.110.063025",
    journal = "Phys. Rev. D",
    volume = "110",
    number = "6",
    pages = "063025",
    year = "2024"
}

@article{Yang_2024,
   title={Searching accretion-enhanced dark matter annihilation signals in the Galactic Centre},
   volume={2024},
   ISSN={1029-8479},
   url={http://dx.doi.org/10.1007/JHEP10(2024)094},
   DOI={10.1007/jhep10(2024)094},
   number={10},
   journal={Journal of High Energy Physics},
   publisher={Springer Science and Business Media LLC},
   author={Yang, Meiwen and Guo, Zhi-Qi and Luo, Xiao-Yi and Shen, Zhao-Qiang and Xia, Zi-Qing and Lu, Chih-Ting and Tsai, Yue-Lin Sming and Fan, Yi-Zhong},
   year={2024},
   month=Oct }

@article{Dror:2023fyd,
    author = "Dror, Jeff A. and Gori, Stefania and Munbodh, Pankaj",
    title = "{QCD axion-mediated dark matter}",
    eprint = "2306.03145",
    archivePrefix = "arXiv",
    primaryClass = "hep-ph",
    doi = "10.1007/JHEP09(2023)128",
    journal = "JHEP",
    volume = "09",
    pages = "128",
    year = "2023"
}

@article{Armando:2023zwz,
    author = "Armando, Giovanni and Panci, Paolo and Weiss, Joachim and Ziegler, Robert",
    title = "{Leptonic ALP portal to the dark sector}",
    eprint = "2310.05827",
    archivePrefix = "arXiv",
    primaryClass = "hep-ph",
    reportNumber = "TTP23-045, P3H-23-071, MPP-2023-240, FR-PHENO-2023-11",
    doi = "10.1103/PhysRevD.109.055029",
    journal = "Phys. Rev. D",
    volume = "109",
    number = "5",
    pages = "055029",
    year = "2024"
}

@article{Flores:2024hts,
    author = "Flores, C{\'e}sar V. and Lenzi, C. H. and Dutra, M. and Louren{\c{c}}o, O. and Arba{\~n}il, Jos{\'e} D. V.",
    title = "{Gravitational wave asteroseismology of dark matter hadronic stars}",
    eprint = "2402.12600",
    archivePrefix = "arXiv",
    primaryClass = "hep-ph",
    doi = "10.1103/PhysRevD.109.083021",
    journal = "Phys. Rev. D",
    volume = "109",
    number = "8",
    pages = "083021",
    year = "2024"
}

@article{Sen:2024yim,
    author = "Sen, Debashree and Guha, Atanu",
    title = "{Impact of a dark boson mediated feeble interaction between dark matter and hadronic matter on the f-mode oscillation of neutron stars}",
    eprint = "2409.18890",
    archivePrefix = "arXiv",
    primaryClass = "hep-ph",
    doi = "10.1103/PhysRevD.110.103013",
    journal = "Phys. Rev. D",
    volume = "110",
    number = "10",
    pages = "103013",
    year = "2024"
}

@article{Tsang:2020,
    author = "Tsang, C. Y. and Tsang, M. B. and Danielewicz, Pawel and Lynch, W. G. and Fattoyev, F. J.",
    title = "{Impact of the neutron-star deformability on equation of state parameters}",
    doi = "10.1103/PhysRevC.102.045808",
    journal = "Phys. Rev. C",
    volume = "102",
    number = "4",
    pages = "045808",
    year = "2020"
}

@article{Keller2023,
    author = "Keller, J. and Hebeler, K. and Schwenk, A.",
    title = "{Nuclear Equation of State for Arbitrary Proton Fraction and Temperature Based on Chiral Effective Field Theory and a Gaussian Process Emulator}",
    doi = "10.1103/PhysRevLett.130.072701",
    journal = "Phys. Rev. Lett.",
    volume = "130",
    number = "7",
    pages = "072701",
    year = "2023"
}

@article{Thakur_2025,
   title={Feasibility study of a dark matter admixed neutron star based on recent observational constraints},
   volume={697},
   ISSN={1432-0746},
   url={http://dx.doi.org/10.1051/0004-6361/202451879},
   DOI={10.1051/0004-6361/202451879},
   journal={Astronomy \& Astrophysics},
   publisher={EDP Sciences},
   author={Thakur, P. and Malik, T. and Das, A. and Jha, T. K. and Sharma, B. K. and Providência, C.},
   year={2025},
   month=May, pages={A220} }

@article{Karkevandi_2024,
    author = "Karkevandi, Davood Rafiei and Shahrbaf, Mahboubeh and Shakeri, Soroush and Typel, Stefan",
    title = "{Exploring the Distribution and Impact of Bosonic Dark Matter in Neutron Stars}",
    doi = "10.3390/particles7010011",
    journal = "Particles",
    volume = "7",
    number = "1",
    pages = "201--213",
    year = "2024"
}

@Article{Grippa2025,
AUTHOR = {Grippa, Francesco and Lambiase, Gaetano and Poddar, Tanmay Kumar},
TITLE = {Searching for New Physics in an Ultradense Environment: A Review on Dark Matter Admixed Neutron Stars},
JOURNAL = {Universe},
VOLUME = {11},
YEAR = {2025},
NUMBER = {3},
ARTICLE-NUMBER = {74},
URL = {https://www.mdpi.com/2218-1997/11/3/74},
ISSN = {2218-1997},
DOI = {10.3390/universe11030074}
}

@article{Mauviard:2025,
   title={A {NICER} View of the 1.4 {M}$_\odot$ {Edge-on Pulsar PSR J0614-3329}},
   volume={995},
   ISSN={1538-4357},
   url={http://dx.doi.org/10.3847/1538-4357/ae145d},
   DOI={10.3847/1538-4357/ae145d},
   number={1},
   journal={The Astrophysical Journal},
   publisher={American Astronomical Society},
   author={Mauviard, Lucien and Guillot, Sebastien and Salmi, Tuomo and Choudhury, Devarshi and Dorsman, Bas and González-Caniulef, Denis and Hoogkamer, Mariska and Huppenkothen, Daniela and Kazantsev, Christine and Kini, Yves and Olive, Jean-Francois and Stammler, Pierre and Watts, Anna L. and Mendes, Melissa and Rutherford, Nathan and Schwenk, Achim and Svensson, Isak and Bogdanov, Slavko and Kerr, Matthew and Ray, Paul S. and Guillemot, Lucas and Cognard, Ismaël and Theureau, Gilles},
   year={2025},
   month=dec, pages={60} }

@article{Andersson:1997rn,
    author = "Andersson, Nils and Kokkotas, Kostas D.",
    title = "{Towards gravitational wave asteroseismology}",
    eprint = "gr-qc/9711088",
    archivePrefix = "arXiv",
    doi = "10.1046/j.1365-8711.1998.01840.x",
    journal = "Mon. Not. Roy. Astron. Soc.",
    volume = "299",
    pages = "1059--1068",
    year = "1998"
}

@article{Thakur:2025fmode,
    author = "Thakur, Prashant and Rather, Ishfaq Ahmad and Lim, Y.",
    title = "{Effect of dark matter and $\sigma$-cut potential on radial and nonradial oscillation modes in neutron stars}",
    eprint = "2507.13227",
    archivePrefix = "arXiv",
    doi = "10.1103/7tnx-s66h",
    journal = "Phys. Rev. D",
    volume = "112",
    pages = "043017",
    year = "2025"
}

@article{Chandrakar:2026fmode,
    author = "Chandrakar, Harsh and Taridalu, Aravind and Thakur, Prashant and Malik, Tuhin and Jha, T. K.",
    title = "{Non-radial oscillations in dark matter admixed neutron stars using a two fluid formalism}",
    doi = "10.1016/j.jspc.2026.100304",
    journal = "J. Subatomic Part. Cosmol.",
    volume = "5",
    pages = "100304",
    year = "2026"
}

@article{Shahrbaf:2026fmode,
    author = "Shahrbaf, Mahboubeh and Thakur, Prashant and Rafiei Karkevandi, Davood",
    title = "{Probing strange dark matter through f-mode oscillations of neutron stars with hyperons and quark matter}",
    eprint = "2510.08115",
    archivePrefix = "arXiv",
    doi = "10.1088/1475-7516/2026/03/017",
    journal = "JCAP",
    volume = "03",
    pages = "017",
    year = "2026"
}

@article{Rather:2025fmode,
    author = "Rather, Ishfaq Ahmad and Marquez, Kau D. and Thakur, Prashant and Louren\c{c}o, Odilon",
    title = "{Nonradial oscillation modes in hybrid stars with hyperons and delta baryons}",
    eprint = "2412.12002",
    archivePrefix = "arXiv",
    doi = "10.1103/7qns-616m",
    journal = "Phys. Rev. D",
    volume = "112",
    pages = "023013",
    year = "2025"
}

@article{LZ:2019sgr,
    author = "Akerib, D. S. and others",
    collaboration = "LZ",
    title = "{The LUX-ZEPLIN (LZ) Experiment}",
    eprint = "1910.09124",
    archivePrefix = "arXiv",
    primaryClass = "physics.ins-det",
    reportNumber = "FERMILAB-PUB-19-555-AE-E",
    doi = "10.1016/j.nima.2019.163047",
    journal = "Nucl. Instrum. Meth. A",
    volume = "953",
    pages = "163047",
    year = "2020"
}

@article{XENON:2024wpa,
    author = "Aprile, E. and others",
    collaboration = "XENON",
    title = "{The XENONnT dark matter experiment}",
    eprint = "2402.10446",
    archivePrefix = "arXiv",
    primaryClass = "physics.ins-det",
    doi = "10.1140/epjc/s10052-024-12982-5",
    journal = "Eur. Phys. J. C",
    volume = "84",
    number = "8",
    pages = "784",
    year = "2024"
}

@article{Fornal:2018eol,
    author = "Fornal, Bartosz and Grinstein, Benjamin",
    title = "{Dark Matter Interpretation of the Neutron Decay Anomaly}",
    eprint = "1801.01124",
    archivePrefix = "arXiv",
    primaryClass = "hep-ph",
    doi = "10.1103/PhysRevLett.120.191801",
    journal = "Phys. Rev. Lett.",
    volume = "120",
    number = "19",
    pages = "191801",
    year = "2018",
    note = "[Erratum: Phys.Rev.Lett. 124, 219901 (2020)]"
}

@article{Issifu:2025jac,
    author = "Issifu, Adamu and Thakur, Prashant and Rafiei Karkevandi, Davood and da Silva, Franciele M. and Menezes, D{\'e}bora P. and Lim, Y. and Frederico, Tobias",
    title = "{Dark matter heating in evolving protoneutron stars: A two-fluid approach}",
    eprint = "2511.07567",
    archivePrefix = "arXiv",
    primaryClass = "astro-ph.HE",
    doi = "10.1103/bhm3-jzjq",
    journal = "Phys. Rev. D",
    volume = "113",
    number = "10",
    pages = "103042",
    year = "2026"
}
\end{document}